\documentclass[preprint,12pt]{elsarticle}

\usepackage{amsmath,amsthm,amsfonts,amssymb,amscd,amstext}
\newcommand{\ppf}{P_{\operatorname{ppf}}}
\usepackage{bm} %makes argument bold
\usepackage{siunitx} % For units, such as angles or nano meters
\usepackage{color}
\makeatletter
\def\ps@pprintTitle{%
 \let\@oddhead\@empty
 \let\@evenhead\@empty
 \def\@oddfoot{\centerline{\thepage}}%
 \let\@evenfoot\@oddfoot}
\makeatother
\begin{document}

\begin{frontmatter}

%% Title, authors and addresses

%% use the tnoteref command within \title for footnotes;
%% use the tnotetext command for theassociated footnote;
%% use the fnref command within \author or \address for footnotes;
%% use the fntext command for theassociated footnote;
%% use the corref command within \author for corresponding author footnotes;
%% use the cortext command for theassociated footnote;
%% use the ead command for the email address,
%% and the form \ead[url] for the home page:
%% \title{Title\tnoteref{label1}}
%% \tnotetext[label1]{}
%% \author{Name\corref{cor1}\fnref{label2}}
%% \ead{email address}
%% \ead[url]{home page}
%% \fntext[label2]{}
%% \cortext[cor1]{}
%% \affiliation{organization={},
%%             addressline={},
%%             city={},
%%             postcode={},
%%             state={},
%%             country={}}
%% \fntext[label3]{}

\title{Retrieval of aerosol properties from in situ, multi-angle light scattering measurements using invertible neural networks}

%% use optional labels to link authors explicitly to addresses:
%% \author[label1,label2]{}
%% \affiliation[label1]{organization={},
%%             addressline={},
%%             city={},
%%             postcode={},
%%             state={},
%%             country={}}
%%
%% \affiliation[label2]{organization={},
%%             addressline={},
%%             city={},
%%             postcode={},
%%             state={},
%%             country={}}

\author[inst1,inst3]{Romana Boiger}
\author[inst2,inst3]{Rob L. Modini}
\author[inst2]{Alireza Moallemi}
\author[inst1]{David Degen}
\author[inst2]{Martin Gysel-Beer}
\author[inst1,inst4]{Andreas Adelmann}

\affiliation[inst1]{organization={Paul Scherrer Institut, Laboratory for Scientific Computing and Modelling},%Department and Organization
            addressline={Forschungsstrasse 111}, 
            city={Villigen},
            postcode={5232}, 
            country={Switzerland}}
\affiliation[inst2]{organization={Paul Scherrer Institut, Laboratory of Atmospheric Chemistry},%Department and Organization
            addressline={Forschungsstrasse 111}, 
            city={Villigen},
            postcode={5232}, 
            country={Switzerland}}           
\affiliation[inst3]{organization={shared first authors}}
\affiliation[inst4]{organization={corresponding author, andreas.adelmann@psi.ch}}
\begin{abstract}
%% Text of abstract
Atmospheric aerosols have a major influence on the earth's climate and public health. Hence, studying their properties and recovering them from light scattering measurements is of great importance. State of the art retrieval methods such as pre-computed look-up tables and iterative, physics-based algorithms can suffer from either accuracy or speed limitations. These limitations are becoming increasingly restrictive as instrumentation technology advances and measurement complexity increases. Machine learning algorithms offer new opportunities to overcome these problems, by being quick and precise. In this work we present a method, using invertible neural networks to retrieve aerosol properties from in situ light scattering measurements. In addition, the algorithm is capable of simulating the forward direction, from aerosol properties to measurement data. The applicability and performance of the algorithm are demonstrated with simulated measurement data, mimicking in situ laboratory and field measurements. With a retrieval time in the millisecond range and a weighted mean absolute percentage error of less than 1.5\%, the algorithm turned out to be fast and accurate. By introducing Gaussian noise to the data, we further demonstrate that the method is robust with respect to measurement errors. In addition, realistic case studies are performed to demonstrate that the algorithm performs well even with missing measurement data. 

\end{abstract}

%%Graphical abstract
%\begin{graphicalabstract}
%\includegraphics{grabs}
%\end{graphicalabstract}

%%Research highlights
%\begin{highlights}
%\item Research highlight 1
%\item Research highlight 2
%\end{highlights}

\begin{keyword}
%% keywords here, in the form: keyword \sep keyword
deep neural networks \sep aerosol property retrieval \sep in situ scattering measurements \sep inverse modelling
%% PACS codes here, in the form: \PACS code \sep code
\PACS 0000 \sep 1111
%% MSC codes here, in the form: \MSC code \sep code
%% or \MSC[2008] code \sep code (2000 is the default)
\MSC 0000 \sep 1111
\end{keyword}

\end{frontmatter}

%% \linenumbers

%% main text
\section{Introduction/Motivation}
Atmospheric aerosols are small particles suspended ubiquitously throughout the Earth's atmosphere. These particles have important impacts on the Earth's climate \cite{Myhre_2013} and public health \cite{Burnett_2018}, which are governed by aerosol properties such as concentration, size, and composition. These properties can change drastically in time and space. A conceptually simple but powerful method for characterizing the variable properties of an aerosol sample is to measure the angular distribution of light that is scattered from it. Such multi-angle light scattering measurements can be performed both in situ (aerosol sample drawn from the atmosphere into an instrument known as a polar nephelometer; e.g. \cite{Barkey_2012}) and remotely (using satellite, aircraft-borne, or ground-based sensors, e.g. \cite{Dubovik2019}). The resulting measurements contain substantial information about the concentration, size, shape, and complex refractive index of the aerosol being probed. Inversion methods are then required to retrieve this information.

The most sophisticated inversion methods rely on iterative optimization algorithms that solve forward models describing the underlying physics of light scattering by gas molecules and small particles \cite{Bohren1998}. In the simplified case when the particles are assumed to be spherically shaped, the forward models are based on Mie theory, which provides an analytical solution to Maxwell’s equations for a plane monochromatic wave incident on a homogeneous sphere of arbitrary radius \cite{Mie1908}. For the more realistic case of arbitrarily shaped aerosol particles, more complex and computationally expensive superposition-based forward models are required such as the multi-sphere T-matrix method \cite{Mishchenko_1996}, or those based on the Discrete Dipole Approximation \cite{Draine_1994}. Efficient and optimized computer codes that implement these types of forward models (e.g. \cite{Mishchenko_2000, Draine_2013,gmd-11-2739-2018}), as well as their use in iterative inversion schemes (e.g. GRASP-OPEN \cite{Dubovik_2011}) are now freely available. 

The main limitation of iterative, physics-based retrieval algorithms is that they are too computationally expensive and slow for some important applications. In the remote sensing context, such applications include the operational processing in near real-time or reanalysis of large volumes of satellite data for downstream use in climate, air quality and weather models (e.g. \cite{Benedetti_2009}). In the in situ context, such applications include the near real-time processing of polar nephelometer measurements to retrieve particular aerosol properties, which is becoming increasingly important with the development of distributed sensor networks, and the corresponding drive to miniaturize instruments (e.g. \cite{Chen_2020}). 

The traditional approach for solving the problem of retrieval speed is to use pre-computed look-up tables (LUTs). LUTs contain limited sets of simulated multi-angle light scattering signals that correspond to discrete sets of combinations of the parameters to be retrieved. A retrieval is performed by selecting the set of parameters whose simulated signals most closely matches a particular measurement. The LUT approach is currently the most widely used method for the operational processing of satellite-obtained aerosol remote sensing data \cite{Dubovik2019}. However, although LUTs are able to solve the problem of retrieval speed, they do so at the cost of retrieval accuracy, since they necessarily involve parameter selectivity and discretization. This problem is becoming even more of a limitation with the continued development of advanced polarimetric light scattering measurements (remote sensing and in situ) \cite{Dolgos_2014, Dubovik2019}, which have higher information content concerning a larger number of free parameters, in particular also for combined multi-parameter retrievals using data from multiple sensors. E.g. combined surface reflectivity and aerosol property retrieval.

Machine learning algorithms present new opportunities for solving the aerosol property retrieval problem with sufficient speed and accuracy to overcome some of the limitations of previous approaches. Some very early efforts in this direction began already in the 1990s and involved the application of neural networks to different forms of light scattering data \cite{Ishimaru_1990, Wang_1999, Ulanowski_1998, Berdnik_2004, Berdnik_2009}. Since then, much work has focused on remotely sensed light scattering data, as reviewed recently in \cite{Di_noia_2018}. In the remote sensing context, it seems that a combination of tools (e.g. neural networks as providers of initial guesses or as forward models in iterative optimization schemes) may ultimately prove to be the most effective way of operationally processing the large amounts of data collected by satellite-borne and ground-based sensors \cite{Fan_2019, Shi_2020, Di_noia_2015, Di_noia_2017}.  

In the present study, we return to the comparatively less-studied problem of the inversion of in situ light scattering phase function measurements using neural networks. Although similar in principle to remote sensing measurements, in situ measurements are unique in that the single scattering approximation is typically valid (i.e., additional light scattering between the sensitive volume and detector is negligible.), and there is no need to account for light reflections from different types of Earth surfaces (e.g., water, ice, different land types) as is necessary in satellite or airborne remote sensing. This simpler configuration makes it possible to keep the problem constrained on light scattering phase functions and associated aerosol property retrieval. 

This study has a concrete and practical motivation: to train models that will be applied to measurements obtained with the new polarized, laser-imaging type polar nephelometer \cite{Dolgos_2014} that is currently being developed within the Aerosol Physics Group at the Paul Scherrer Institute (PSI). This instrument will eventually be used in laboratory experiments to measure angular distribution of light scattering $y$ (e.g. phase functions and polarized phase functions at multiple wavelengths) for different types of aerosols described with the properties $x$ (e.g. size distribution parameters, complex refractive index). We present here a proof of concept showing that neural network models can be successfully applied to the problem of quickly and accurately retrieving the aerosol properties $x$ from the light scattering measurements $y$ that will be obtained with this instrument. For this purpose, we use synthetic data sets with realistically simulated measurement uncertainties. In follow up studies, experiments will be performed with the instrument to further evaluate the performance of the neural network based retrievals presented here, as well as to evaluate the performance of common physics-based retrieval algorithms used in aerosol remote sensing (e.g. \cite{Schuster_2019}).            

A key novelty of the present study relative to other recent studies (e.g. \cite{Berdnik_2016, Xu_2021}) is the model architecture that we consider. In particular, we focus on a class of neural network models known as invertible neural networks (INNs) \cite{Ardizzone2018}. A single INN model trained in one direction on a particular data set has the unique feature that it can be run in both the forward (aerosol properties to light scattering data, $x \to y$) and inverse (light scattering data to aerosol properties, $y \to x$) directions with negligible computational cost. This feature creates great flexibility with respect to the model application. 

In Section 2 we present a theoretical overview of the problem and INN model architecture. Section 3 introduces the synthetic data sets we use for model training and validation, while Section 4 discusses the implementation of the model including the data preprocessing steps. In Section 5 we present the results and in Section 6 we discuss the conclusions and outlook of the study.

\section{Theory}
\subsection{Problem description}\label{sec:problem_description}
In this work we retrieve aerosol properties from multi-angle multi-wavelength and polarized light scattering data. Let $x \in X\subseteq \mathbb{R}^N$, denote the aerosol properties, like spectral complex refractive index or particle size distribution parameters, where $N$ is the total number of properties. The functions obtained by the measurement device are the phase function $P_{11}=P_{11}(\theta)$ (i.e. angularly-resolved scattered light intensity) and the polarized phase function $-\frac{P_{12}}{P_{11}}=-\frac{P_{12}(\theta)}{P_{11}(\theta)}$ (i.e. angularly resolved relative degree of linear polarization of scattered light for unpolarized incident light), where $\theta$ is the angle.
These are (normalized) elements of the scattering matrix used in the Stokes formalism, see e.g. \cite{Dolgos_2014, amt-2021-251} for more details.

%These are elements of the scattering matrix $P(\theta)=\begin{pmatrix} P_{11}(\theta) & P_{12}(\theta) & 0 & 0 \\ P_{12}(\theta) & P_{22}(\theta) & 0 & 0 \\ 0 & 0 & P_{33}(\theta) & P_{34}(\theta) \\ 0& 0& -P_{34}(\theta)& P_{44}(\theta)\end{pmatrix}$, where $\theta$ denotes the polar scattering angle, i.e. the angle between the illuminating and scattered light. The scattering matrix itself is used in the Stokes formalism, that describes the scattering process, see e.g. \cite{Dolgos_2014}. The Stokes vector of incident light beam is related with the Stokes vector describing the scattered light per angle via the scattering matrix $P(\theta)$:
%\begin{equation*}
% \begin{pmatrix}I_{scat}\\Q_{scat}\\U_{scat}\\V_{scat}\end{pmatrix}=K P(\theta) \begin{pmatrix}I_{in}\\Q_{in}\\U_{in}\\V_{in}\end{pmatrix}.
%\end{equation*}
%Here, $I$ is the irradiance and $Q$, $U$ and $V$ denote the Stokes parameters. 

The forward problem is now to compute the phase and polarized phase functions $y \in \{P_{11}, -\frac{P_{12}}{P_{11}} \}$ from aerosol properties $x \in X$: $y = F(x)$. Where $F$ indicates the underlying Mie theory or other theories of aerosol light scattering. 

The inverse problem is to retrieve the aerosol properties from the measurement data ($y^{\delta}$) : $\hat{x} = F^{-1}(y^{\delta})$, where the $\hat{}$ symbol indicates that $\hat{x}$ is an estimate of the true values $x$.
Due to the ill-posed nature of inverse problems, i.e. either there exists no solution, or the solution is not unique or small errors in the data $y^{\delta}$ can lead to huge errors in the retrieval of $x$, special solution methods, including e.g. regularization, need to be applied to solve inverse problems. 

\begin{figure}
    \centering
    \includegraphics[width =\textwidth]{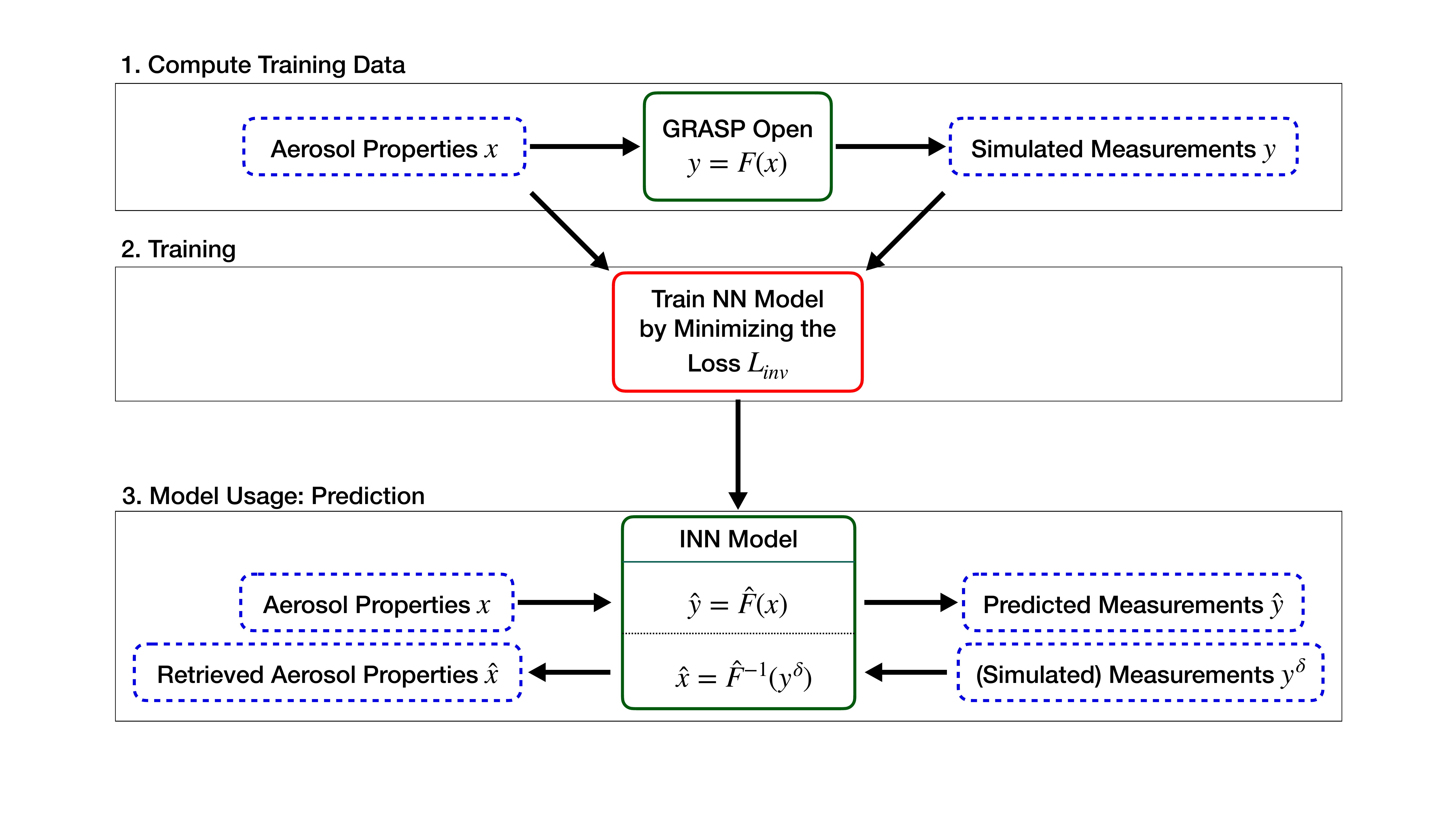}
    \caption{Concept of how to build and use the INN model. To use the invertible neural network model in the inverse direction, a best-of-$n$ strategy is applied, so for a given measurement, $n$ sets of aerosol properties are retrieved, and the best one, according to the forward pass is chosen.}
    \label{fig:model_concept}
\end{figure}

In our work, we focus on so-called invertible neural networks. A detailed description can be found in section \ref{sec:INN}. The main advantage of this method is, that the forward and the inverse problem are solved simultaneously, even though only one model needs to be trained. This is due to the architecture of the INN.
The general concept of building the model is depicted in Figure \ref{fig:model_concept}. Training the INN requires data, which can be obtained either from simulations or measurements. The trained model can then be used to either predict light scattering measurement data from a given set of aerosol properties, or to retrieve a set of aerosol properties from measured or simulated light scattering data. One key advantage of having forward and inverse pass:
it is possible to assess how well the retrieved simplified surrogate aerosol model represents the measured phase functions (even angularly resolved if needed).

%\subsection{Measurement errors}
%\textcolor{red}{Do we really want to include measurement errors in %this paper? We do not consider it in the INN models}

\subsection{Invertible neural networks}
\label{sec:INN}
The used invertible neural network was first introduced by Ardizzone et al. \cite{Ardizzone2018}, and is described here for the benefit of readability. Input and output of the network are divided randomly into two halves, $x=(x_1,x_2)$, $y = (y_1,y_2)$. The main components of the INN are so called, affine coupling blocks, having the following simplified structure
\begin{align*}
    y_1 &= x_1 \odot \exp{(s \cdot \arctan (\mathrm{NN}(x_2)))}+\mathrm{NN}(x_2) \\
    y_2 &= x_2 \odot \exp{(s \cdot \arctan (\mathrm{NN}(y_1)))}+\mathrm{NN}(y_1) \\
    x_1 &= (y_1-\mathrm{NN}(x_2)) \odot \exp{(-s \cdot \arctan(\mathrm{NN}(x_2)))} \\
    x_2 &= (y_2-\mathrm{NN}(x_1))\odot \exp{(-s \cdot \arctan (\mathrm{NN}(y_1)))}, 
\end{align*}
where $s$ is a scaling variable, that damps, together with $\arctan$, the exponential function,  and $\mathrm{NN}$ stands for an arbitrary neural network. In theory, four different neural networks could be used, but this would increase the number of hyperparameters - so in this work, all $\mathrm{NN}$ are dense and have the same width and depth. Hence it is clear, that $x_1, x_2, y_1$ and  $y_2$ need to have the same dimension. This is guaranteed by padding both input and output by zero or low noise, $x_{pad} \in \mathbb{R}^{d_{P_x}}$, $y_{pad} \in \mathbb{R}^{d_{P_y}}$, where $d_{P_x}$ and $d_{P_y}$ are the corresponding dimensions, that can be zero as well. With this choice, the invertibility of the INN is also assured. 
The INN is then given by a concatenation of affine coupling blocks and permutation layers. The permutation layers guarantee that the splitting of input and output into two parts is not always the same. To capture the information about $x$ that is not contained in the measurements $y \in \mathbb{R}^{2M}$, a latent output variable $z \in \mathbb{R}^{d_Z}$, with dimension $d_Z$, 
is introduced. The dimension $M$ for the measurements denotes the number of measured angles, so it varies between 2 and 178. During the inverse pass, these latent variables are sampled from a standard normal distribution.
The forward pass of the INN, $\hat{F}$, is then given by:

\begin{align*}\hat{F}:\mathbb{R}^N \times \mathbb{R}^{d_{P_x}} & \rightarrow \mathbb{R}^{2M}\times \mathbb{R}^{d_Z}\times \mathbb{R}^{d_{P_y}} \\ 
(x, x_{pad} )&\mapsto (\hat{y}, z, y_{pad}) \\ 
\text{s.t. } \hat{y} &\approx f(x)
\end{align*}
and the inverse pass, $\hat{F}^{-1}$, is:
\begin{align*}\hat{F}^{-1}:\mathbb{R}^{2M}\times \mathbb{R}^{d_Z}\times \mathbb{R}^{d_{P_y}} &\rightarrow \mathbb{R}^N \times \mathbb{R}^{d_{P_x}} \\
(y,z, y_{pad}) &\mapsto (\hat{x}, x_{pad}) \\
\text{s.t. } \hat{x} &\approx f^{-1}(y)
\end{align*}
and it holds that $\hat{F}(\hat{F}^{-1}(y,z,y_{pad}))= (y,z,y_{pad})$ and $\hat{F}^{-1} (\hat{F}(x,x_{pad}))= (x,x_{pad})$.
To train the neural network a loss function needs to be defined. In this case, it is a composition of numerous loss functions:
\begin{equation*}
    L_{inv} = w_x L_{x} + w_y L_{y}
        + w_z L_{z} + w_r L_{r}
        + w_{p} L_{p}.
\end{equation*}

The loss $L_x=\|p_x(\hat{x})-p_x(x)\|^2$ ensures that the sampled aerosol property distributions, $p_x(x)$,  match the distributions of the aerosol properties in the data set, $p_x(\hat{x})$. Likewise, the loss $L_z= \|p_z(z|\hat{y})-p_z(z) \|^2$ assures that the latent variable is sampled from the desired normal distribution. The losses $L_y=  \sum^{N}_{i=1}{\|\hat{F}(x_i)-y_i\|^2}$ and $L_r = \sum^{N}_{i=1}{\|\hat{F}^{-1}(\hat{F}(x_i)+\varepsilon)-x_i\|^2} $ make sure that the forward and inverse predictions, resp., mimic the data. Through the introduction of Gaussian noise $\varepsilon \sim \mathcal{N}(0, \sigma_r)^{2M+d_z+d_{P_y}}$ in $L_r$, robustness of the inverse prediction should be guaranteed. Lastly, the loss $L_p= \|x_{pad} \|^2 + \|y_{pad}\|^2$ ensures that the amplitude of the noise fed into the model through the padding dimensions is low, so that no information is encoded there. 
Further details on INNs can be found in the work of Ardizzone et al., \cite{Ardizzone2018}.

The model is trained w.r.t the forward pass, meaning that the input consists of a set of aerosol properties and that the output, or predictions, consists of the phase and polarized phase functions. Due to the structure of the INN, the inverse pass comes with no additional effort, meaning, the network does not need to be trained in addition with the (polarized) phase functions as input and the aerosol properties as output. Although, of course, the loss function contains forward and inverse prediction errors. 
As discussed in Section \ref{sec:problem_description}, we are faced with the challenge that the inverse direction is not unique, such that several sets of aerosol properties can lead to similar phase and polarized phase functions. The INN architecture considers this problem on the one hand by the special loss function and on the other hand by the introduction of the latent space. Therefore, for the retrieval of aerosol properties, we apply a best-of-$n$ strategy, meaning that for a given measurement of angular resolved scattering data $n$ aerosol properties are predicted. The variation of the measurements to get the $n$ predictions is done via the latent space.  All of these $n$ properties are then handed back to the forward pass, and the set of aerosol properties whose associated phase functions are closest to the given data are chosen to be the predicted aerosol properties.

\section{Synthetic data sets}
The GRASP-OPEN forward model \cite{Dubovik_2011} is used to generate the synthetic data to train the INN models. 
In this work, two types of data, corresponding to two different applications of in situ multi-angle light scattering measurements are considered. These applications are chosen because they mimic the type of measurements that will be performed with the PSI polar nephelometer. 
\begin{enumerate}
    \item Simulation of aerosol measurements in a laboratory: specifically, measurements of spherical, monodisperse, pure-component aerosols. Such aerosols can be routinely generated in the laboratory using common aerosol generation techniques (e.g. nebulization of aqueous solutions of known composition) combined with size classification (e.g. by particle electrical mobility or by particle aerodynamic diameter). 
    \item Simulation of aerosol measurements in the field. This represents the situation of taking the instrument into the field to measure multi-component, ambient aerosols. In this case bi-lognormal modes are chosen to represent atmospheric aerosol size distributions over the diameter range from $\sim$ 50 nm up to $\sim$ 10 µm, with different refractive indices in each mode and allowance for non-spherical particles in the upper-most mode (defined as the coarse mode, see further discussion below in \ref{space_variable_x_field}). 
\end{enumerate}
	
%a.	Pure component aerosols (i.e., 1 set of complex RIs)
%b.	Two externally mixed aerosols with same size distribution but different complex RIs (not so ‘experimentally-relevant’, but will be used for an interesting test of INN retrieval capability, see below)

%2)	Atmospheric aerosols measured in the field. Two different ways to represent size distributions of such aerosols:

%a.	Modal representation. E.g. bi-lognormal modes are commonly chosen to represent atmospheric aerosol size distributions over the diameter range from $\sim$ 50 nm up to $\sim$ 10 µm

%b.	Sectional representation. A fixed number of discrete size bins over the same size range. Typically, 10 – 20 bins are used in aerosol models. 

\subsection{Measurement space variables $y$: virtual polar nephelometer instrument configurations} \label{space_variable_y}

The simulated measurements are chosen such that they match the measurement output of the PSI polar nephelometer. The instrument is designed to measure both scattering phase functions, $P_{11}$ and polarized phase functions, $-\frac{P_{12}}{P_{11}}$ at three different wavelengths $\lambda$ (450, 532, and 630 nm), over the polar scattering angle range from 0° to 179°. Both, $P_{11}$ and $-\frac{P_{12}}{P_{11}}$, will be measured at an angular resolution of approximately 1°.

There are numerous technical challenges associated with polar nephelometry that can result in the loss of measurement information in real experimental setups relative to the ideal measurement configuration. We consider the following real-world artefacts in our simulations:
\begin{enumerate}
    \item Scattered light truncation. Due to physical design limitations, nephelometers can not perform scattered light measurements at extreme forward and backward angles, beyond the so-called truncation angles (e.g. \cite{Moosmuller_2003}). We consider forward and backward truncation angles of 0°-5° and 175°-179°. In addition, we also consider the situation where measurements can not be performed over the range from 85° to 95°, since these angles are associated with higher measurement uncertainties in the PSI polar nephelometer.
    \item Loss of information on polarization dependence (i.e., only $P_{11}$ is measured)
    \item Loss of spectral information (i.e., measurements only performed at 1 or 2 wavelengths). 
\end{enumerate}

To investigate the impact of these practical artefacts on INN model performance we study 22 different cases as outlined in Table \ref{tab:case_study}. First, we simulate the ideal case of having available $P_{11}$ and $-\frac{P_{12}}{P_{11}}$ at all angles from 0° to 179° and with three wavelengths. (Note that since $-\frac{P_{12}}{P_{11}}$ is always zero at the angle 0°, this point is ignored for training, validation and testing even in this ideal case). Then, 21 additional cases are examined that only take into account angles from 5° to 85° and 95° to 175° to better represent actual measurements. These 21 cases correspond to different polarization and spectral combinations: i.e., the different combinations of either $P_{11}$ and/or $-\frac{P_{12}}{P_{11}}$, measured with one, two or three wavelengths.  

%In summary, we will simulate $P_{11}$ and $-\frac{P_{12}}{P_{11}}$ measurement data from two different virtual instrument configurations: 1-$\lambda$ and 3-$\lambda$ versions. In both cases, the missing angles and measurement uncertainties will be the same. For the 1-$\lambda$ configuration, the measurement space will consist of 2×160=320 variables (representing $P_{11}$ and $-\frac{P_{12}}{P_{11}}$ measured 160 angles). For the 3-$\lambda$ configuration, there will be three times as many variables (i.e., 960).  

To have a baseline for the quality of the models, results are compared to the measurement device errors. These errors are not yet fully characterized for the PSI polar nephelometer. Therefore we choose values that slightly overestimate the errors reported for the instruments similiar to the PSI polar nephelometer \cite{Dolgos_2014}. Specifically, we consider the relative error in $P_{11}$ to be $5\%$ and the absolute error in $-\frac{P_{12}}{P_{11}}$ to be $0.1$.

%A single-set of values chosen from previous literature on the precursor instrument to the PSI polar nephelometer \cite{Dolgos_2014}. $5\% $ relative error in $P_{11}$ $0.1$ absolute error in $-\frac{P_{12}}{P_{11}}$.

\subsection{State space variables $x$: describing spherical, monodisperse aerosols measured in the laboratory}

In the first data set we assume the aerosol particles are homogeneous spheres for three reasons: i) Mie theory is applicable, ii) spherical particles can be generated easily in the laboratory (e.g. polystyrene latex spheres, organic aerosols) and iii) it is a common assumption for retrieving fine mode properties from polarimetric data.

As described in Section \ref{sec:problem_description}, phase and polarized phase function depend on the aerosol size and shape distributions and the complex refractive index. 
The volume size distributions of monomodal aerosols $V(r)$ (where $r$ represents aerosol particle radius) can be represented by lognormal functions, which can be described by three parameters: the total volume concentration, $V_{tot}$, the mean radius $R_{mean}$ and the geometric standard deviation $\sigma_g$:
\begin{equation}\label{size_distribution}
    V(r) = \sum^{N}_{i=1}{\frac{V_{tot}}{\sqrt{2 \pi}\ln(\sigma_{g,i})}}\exp{\left(\frac{-(\ln(r)-\ln(R_{mean,i}))^2))}{2(\ln(\sigma_{g,i}))^2}\right)},
\end{equation}
where $N=1$ for the monomodal case and $N=2$ for the bimodal case (described in Section \ref{space_variable_x_field}).

The complex refractive index is wavelength-dependent and contains two parts: real and imaginary, i.e., $m_{\lambda}=n_{\lambda}+ik_{\lambda}$.

%-	Represents the case where a polar nephelometer is used to measure the scattering phase function of size-classified, spherical aerosols. This is one of the simplest use-cases of the instrument. 

%-	Eventually, such measurements will be performed with the PSI polar nephelometer to validate the  modelling results, see results section X.X below. Thus, parameter ranges below are chosen to cover an experimentally-achievable range of parameter space 

%-	Scattering phase functions depend on the aerosol size and shape distributions, and complex refractive index:

%o	For now we assume only spherical particles for simplicity (Mie theory applicable) and because spherical particles can be easily generated in the laboratory (e.g. PSL spheres, DOS, SOA)  

%Considering the above, the state space for this type of data set will consist of a maximum of $3j+2lj+n$ variables, where $j$ represents the number of size distribution modes (1 in this case), $l$ represents the number of optical wavelengths the measurements/simulations are performed at ($l=1$ or 3 in this case) and $n$ is the number of shape parameters (0 in this case). 

Typically for atmospheric aerosols, variations in refractive index values are minor, or at least smooth, over the range of visible wavelengths that we consider here. To incorporate this knowledge into our simulated data we use the following relationships, which can be considered as approximately valid for aerosols over the $\lambda$ range from 450 to 630 nm. For the real part of the refractive index we assume $n_{\lambda_2}= n_{\lambda_1}$, for two different wavelengths $\lambda_1$, $\lambda_2$. For the complex part of the refractive index we assume
$k_{\lambda_2} \approx k_{\lambda_1}(\lambda_2/\lambda_1)^{1-AAE}$. Here $AAE$ is an often reported parameter known as the absorption Angstrom exponent, which is defined through the empirical relationship $b_{abs, \lambda_1}/b_{abs, \lambda_2} \approx (\lambda_2/\lambda_1)^{-AAE}$ (with the relationship to $k$ then given by $b_{abs, \lambda} \propto C_N k/ \lambda$, which is valid for fixed particle size and $n$, and where $C_N$ represents the particle number density \cite{Moosmuller_2009, Laskin_2015}). The parameter $b_{abs, \lambda}$ is known as the aerosol light absorption coefficient, and it is easily measurable. Therefore, $AAE$ values are reported widely in the aerosol literature. For black carbon aerosols (i.e., highly absorbing carbonaceous aerosols) with small particle size, $AAE \approx 1$ over the visible range, which is a consequence of $k$ being approximately wavelength-independent across visible wavelengths for such aerosols. For brown carbon (i.e. moderately absorbing carbonaceous aerosols) $AAE>1$, which is a consequence of $k$ decreasing with increasing $\lambda$ across visible wavelengths. 
Given these relationships, simulating measurements for the 3-$\lambda$ version of the instrument only requires one extra state space variable – $AAE$ – than is required for the 1-$\lambda$ simulations.

Summarized, the state space consists of at most $4+l$ variables, where $l$ is the number of wavelengths. The variables, together with their lower and upper bounds are listed in Table \ref{tab:monomodal_variables}.

\begin{table}[]
\begin{tabular}{lrr}
\textbf{Variables }& \textbf{Lower bound} & \textbf{Upper bound} \\
$V_{tot}$ [µm$^3$ cm$^{-3}$]         & 1                    & 5000        \\
$R_{mean}$ [nm]                  & 150                  & 2500                 \\
$\sigma_g$ [ ]                        & 1.4                  & 1.45                 \\
$n$ [ ]                      & 1.33                 & 1.6                  \\
$k_{450}, k_{532}, k_{630}$ [ ]\footnotemark[1]                      & 1e-4                    & 0.2                  \\
$AAE$ [ ]\footnotemark[1]                    & 1                    & 7                   
\end{tabular}
\caption{Parameter space covered by the simulations of monomodal, spherical, laboratory-generated aerosols, as defined by the maximum and minimum values for each of the included state space variables.
\newline \tiny{1: As stated in the description $k_{450}, k_{532}, k_{630}$ and $AAE$ are dependent of each other.}}

\label{tab:monomodal_variables}
\end{table}

\subsection{State space variables $x$ describing atmospheric aerosols measured in the field}
\label{space_variable_x_field}

The second data set represents the application of a polar nephelometer to measure ambient atmospheric aerosols in a field setting. This is generally a more complicated and less-controlled use case for the instrument than laboratory use. Atmospheric aerosols are typically complex mixtures of particles of different sizes (covering the diameter range from a few nm’s to 10s of µm), chemical components (i.e., with different complex RIs), and shapes.
Typical simplifications used to represent this complexity:
    \begin{itemize}
        \item Bi-lognormal distributions are used to represent volume size distributions over a broad range, diameters from 50 nm to 10 µm. The smaller size mode peaks at diameters less than 1 µm and is commonly called the ‘fine’ aerosol mode. The larger size mode peaks at diameters greater than 1 µm and is commonly called the ‘coarse’ aerosol mode. In this case, the full size distribution is represented by 6 parameters (3 lognormal parameters for the fine mode, 3 for the coarse mode): Total volume concentration, $V_{tot}$, fine mode fraction, $\chi$, coarse and fine radii, $R_{mean_{fine}}$, $R_{mean_{coarse}}$ and geometric standard deviations, $\sigma_{g_{fine}}$, $\sigma_{g_{coarse}}$. In the present work it is assumed, that the coarse and fine mode mean radii do not overlap. The fine and coarse mode volume concentrations, $V_{fine}$ and $V_{coarse}$ are defined with respect to the total volume concentration $V_{tot}$ and the fine mode fraction, $\chi$: $V_{fine}=\chi  V_{tot}$ and $V_{coarse}=(1-\chi) V_{tot}$. This is, what we use in our case, see equation \ref{size_distribution} with $N=2$ and $i=1$ denotes the coarse and $i=2$ the fine mode.   
        It should be stressed out that the bi-lognormal representation is still a simplified representation of true atmospheric aerosol size distributions. Greater complexity can be considered by representing size distributions with concentration values in discrete size bins (i.e., a sectional representation). Typically, 10 - 22 size bins may be considered. Therefore, sectional representations involve a greater number of aerosol state parameters than modal representations, although smoothness constraints can be applied to limit this number.   
        
        \item In some studies, like Espinosa et al. \cite{Espinosa_2019}, a single, effective complex refractive index (still $\lambda$ dependent) is used to describe the optical properties of the complex atmospheric aerosol as a whole. However, GRASP-OPEN has the ability to simulate the more realistic but still simplified situation where two effective refractive indices are used to describe the aerosol: one for fine mode particles and one for coarse mode particles. We apply this approach here. Also in this bi-modal case, the refractive index values at different wavelengths are strongly related and should not vary independently from each other. Thus, we will again consider the following relationships: $n(\lambda_2)= n(\lambda_1)$ and $k(\lambda_2)= k(\lambda_1)(\lambda_2/\lambda_1)^{1-AAE}$.

        \item Particle non-sphericity is either accounted for with a single parameter (e.g. the fraction of spherical particles in the ensemble, which we denote with the symbol $\phi$), or size-resolved spherical particle fractions (e.g. in up to 22 discrete size bins),  \cite{Dubovik_2006,Dubovik_2011}). We will use a single parameter $\phi$ for our data set. Notice that non-spherical particle fraction is only considered in the coarse mode. All particles in fine mode are assumed to be spherical.
    \end{itemize}

To obtain parameter ranges for simplified bi-modal atmospheric aerosol models we use the comprehensive, airborne polar-nephelometer measurements of Espinosa et al. \cite{Espinosa_2019} performed over the USA. The combined data set can be considered as reasonably representative of the different types of aerosols encountered over continents. They used GRASP-OPEN to retrieve the aerosol parameters (bi-modal size distribution, spherical fraction, complex refractive index) corresponding to the measurements. The results were categorized according to aerosol types (e.g. urban, biogenic, biomass burning, dust-containing aerosols) \footnote{the dust-containing category is termed ‘CO Storms’ in the original paper}. We use the minimum and maximum values obtained across the different aerosol types for the bi-modal size distribution and spherical fraction parameters. However, we use broader ranges for the complex refractive index parameters since only a single effective refractive index was retrieved by these authors, whereas we choose to simulate the more atmospherically relevant case where the fine and coarse mode aerosols each have their own effective refractive index. 
 
Considering the above, the state space for this type of modal data set will consist of a maximum of $11+2j$, where $j$ 
is 0, if only one wavelength is used, otherwise it is 1.  A list of all variables with their lower and upper bounds is given in Table \ref{tab:bimodal_variables}.
%3j+2lj+n variables, where j represents the number of size distribution modes (2 in this case), l represents the number of optical wavelengths the measurements/simulations are performed at (l=1 or 3 in this case) and n is the number of shape parameters (1 in this case). 

%However, it should be considered that for a given aerosol m values are strongly related to each other across the range of visible wavelengths we are considering. Thus, they should not be allowed to vary independently of each other, since this could place us in regions of the parameter space that are not actually accessible to measurement. We will consider the following relationships defined in Section 3.1, which can be considered to be approximately valid for aerosols we will measure over the $\lambda$ range from 450 to 630 nm: 

\begin{table}[]
\begin{tabular}{lrr}
\textbf{Variable (units)}               & \textbf{Lower bound} & \textbf{Upper bound} \\
$V_{tot}$ $[\mu m^3 cm^{-3}]$                           & 9.3                  & 24                   \\
$\chi$                                      & 0.53                 & 0.93                 \\
$R_{mean_{fine}}$ $[\mathrm{\mu m}]$                                & 0.126                & 0.163                \\
$R_{mean_{coarse}}$ $[\mathrm{\mu m}]$                                 & 0.87                 & 1.3                  \\
$\sigma_{g_{fine}}$ [ ]                                     & 1.38                 & 1.57                 \\
$\sigma_{g_{coarse}}$ [ ]                                   & 1.4                 & 1.49                  \\
$n_{fine}$ $[450\,\mathrm{nm}]$                                      & 1.33                 & 1.6                  \\
$k_{fine}$ $[450\,\mathrm{nm}]$                                      & 1e-5                    & 0.2                  \\
$AAE_{fine}$ [ ]                                      & 1                    & 7                    \\
$n_{coarse}$ $[450\,\mathrm{nm}]$                                     & 1.45                 & 1.6                  \\
$k_{coarse}$ $[450\,\mathrm{nm}]$                                      & 0.0001               & 0.02                 \\
$AAE_{coarse}$ [ ]                                      & 1.5                  & 2.4                  \\
$\phi$ [\%] & 17                   & 85                  
\end{tabular}
\caption{Parameter space covered by the simulations of atmospheric aerosols in a bimodal representation, as defined by the maximum and minimum values for each of the included state space variables}
\label{tab:bimodal_variables}
\end{table}

%Sectional representation
%To be assessed after first dealing with the modal-representation data sets described above. 

\subsection{Summary of simulated data sets}
The data sets were created with GRASP-OPEN using the Latin hypercube sampling method based on the parameter spaces stated in Tables \ref{tab:monomodal_variables} and \ref{tab:bimodal_variables}. Both data sets included 100 000 samples. The numbers of measurement and state space variables are summarized in Table \ref{tab:datasets_nr}
.

\begin{table}[]
\begin{tabular}{lrrr}
\textbf{Data set ID} & \textbf{\begin{tabular}[x]{@{}r@{}} Number of measurement  \\ space variables\end{tabular}} & \textbf{\begin{tabular}[x]{@{}r@{}} Number of state \\ space variables\end{tabular}} \\
Lab 1 & 1077 & 5 or 6   \\
Lab 2 & $162 \cdot i \cdot j$ & 5 or 6  \\
Atmos-bimodal 1 & 1077 & 11 or 13   \\
Atmos-bimodal 2 & $162 \cdot i \cdot j$ & 11 or 13  \\
\end{tabular}
\caption{Summary and description of simulated data sets used in this study. "Lab" stands for the case of monomodal, spherical, laboratory-generated aerosols, and "Atmos-bimodal" for the simulations with aerosols in a bimodal representation. The numbers in the Data set ID denote, whether all data are at hand (1), or some data are missing (2). The numbers of measurement space variables include $i = {1,2}$, for either $P_{11}$ or $-\frac{P_{12}}{P_{11}}$ or both and $j= 1$ for one and $j=2$ for two or three wavelengths. The numbers for the state space variables depend on the number of wavelengths.}
\label{tab:datasets_nr}
\end{table}

\section{Implementation and Data Preprocessing}

The models are implemented in Python using the TensorFlow framework and Keras together with the Ray Tune library for the hyperparameter scan. According to the INN architecture, the following hyperparameters need to be chosen: number, depth and width of affine coupling blocks, batch size, learning rate, type of activation function, number of epochs and weights and noise in the loss function. The details are described in section \ref{sec:hyperparameters}. 
All computations were done at the Merlin 6 Cluster at Paul Scherrer Institut using one core. Each cluster node provides \SI{384}{GB} of memory and contains two Intel$^\circledR$ Xeon$^\circledR$ Gold 6152 Processors, with 22 cores and 2 threads per physical core. For the implementations in this work only one core was used.

\subsection{Preprocessing the data}
\begin{figure}
    \centering
    \includegraphics[width=\textwidth]{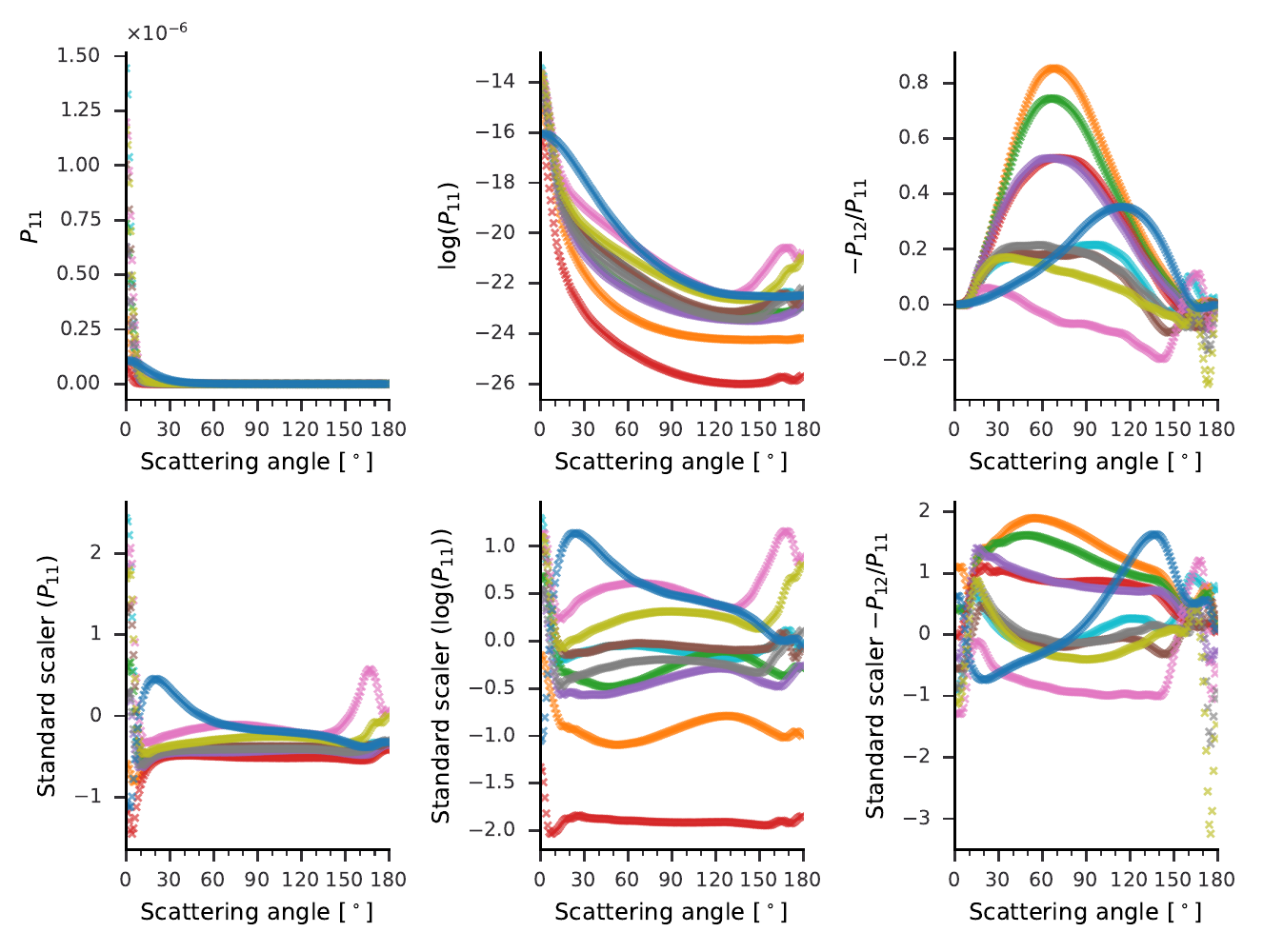}
    \caption{The effects of preprocessing on $P_{11}$ and $-\frac{P_{12}}{P_{11}}$ for a series of randomly chosen spherical, monodisperse aerosols probed with an incident light beam of wavelength \SI{532}{nm}. The left column contains $P_{11}$, the second column $\log(P_{11})$ and the third column $\ppf$. The first row shows the unscaled versions and the second row the application of the StandardScaler.}
    \label{fig:P11_pre}
\end{figure}

Some examples of simulated $P_{11}$ and $-\frac{P_{12}}{P_{11}}$ functions are shown in Fig. \ref{fig:P11_pre}. Each data set consisting of $100\ 000$ of these functions is divided into a training data set and a test data set with a ratio of $80:20$. The training data set in turn is split up into a sub-training data set and a validation data set at each epoch. The validation data set is used to pick the best hyperparameters as discussed in Section \ref{sec:hyperparameters}. Preprocessing is applied to the state space variables $x$ and the measurement space variables $y$ with the goal of achieving a broad spread of the simulated data across all angles. The scikit-learn (\cite{scikit-learn}) {\tt MinMaxScaler} preprocessing function turned out to give good results for the state space variables $x$. The measurement space is divided into the two parts $P_{11}$ and $-\frac{P_{12}}{P_{11}}$, where for $P_{11}$ first the logarithm is applied and then on both parts the scikit-learn {\tt StandardScaler} preprocessing function. As can be seen in Figure~\ref{fig:P11_pre} this preprocessing (second row, second and third column) spreads the data per angle.

\subsection{Hyperparameters} \label{sec:hyperparameters}
An initial random hyperparameter scan was performed for the first model being trained on simulated data: i.e., the model corresponding to the ideal measurement configuration with three wavelengths and all angles. By taking the hyperparameter values from the best model, all but 3 hyperparameters were fixed and used for all the further trained models, also to make the models more comparable amongst each other.  For all models, 3 affine coupling blocks with a depth of 2 and a width of 92 together with a batch size of 8 and a learning rate of $9\times10^{-5}$ were used. The activation functions were chosen to be rectified linear units for the hidden layers and linear for the last layer. The number of epochs was kept at $50$. For the loss function, the artificial weight was kept small, $w_p=0.0005$ and the weight $w_y$ was set to $350$. The noise $\epsilon$ used in the loss $L_r$ was set to $0.1$. 
It turned out, that the remaining weights have a non-negligible influence on the performance of the models, hence a further hyperparameter scan based on grid search was performed for all models. For each weight three values, actually the values of the three best previous hyperparameter scans, where chosen: $w_x \in \{138,142,146\}$, $w_z \in \{291,330,339\}$ and $w_r \in \{258,308,323\}$. Summarized, for the 22 models of the case study, 27 different sets of hyperparameters were chosen to train the corresponding models. Among those models, the best one according to the $R^2$ value of the inverse direction evaluated with the validation data set was selected to be the final model for each case.

%and can be found in Table \ref{tab:hyperparameters}.
%\begin{table}[]
%\begin{tabular}{cc}
%\textbf{Hyperparameter} & \textbf{Values}  \\
%$w_x$ & 138, 142, 146  \\
%$w_z$ & 291, 330, 339 \\
%$w_r$ & 258, 308, 323 \\
%\end{tabular}
%\caption{Hyperparameter values used for scan.}
%\label{tab:hyperparameters}
%\end{table}

\section{Results}

In the following, the results for different cases are presented. Within all models in the hyperparameter scans the best model was chosen according to the highest coefficient of determination $R^2$, scored on the validation data set for the inverse process at the last epoch of training. Notice that this model must not necessarily have the highest coefficient of determination for the forward pass, but the main target of the research is on the inverse model, namely to retrieve aerosol properties. 
To get an idea of the quality of the trained models, the models are further used to make predictions about the previously unseen test data set. Therefore, the relative, absolute and weighted mean absolute percentage errors ($\operatorname{wMAPE}$) are valuable error metrics. Since the absolute error is not suited for comparing results across the parameters, the $\operatorname{wMAPE}$ is used for that. The $\operatorname{wMAPE}$ has the additional advantage over the relative error, that it does not explode when the actual value is zero or very close to zero. The formulae for computing these error metrics are given in the \ref{App:Metrics}.

\subsection{Monomodal case with ideal measurement configuration: $P_{11}$, $-\frac{P_{12}}{P_{11}}$ at all three wavelengths and all angles}
\label{sec:results-monomodal-ideal}
\begin{figure}
    \centering
    \includegraphics[width=\textwidth]{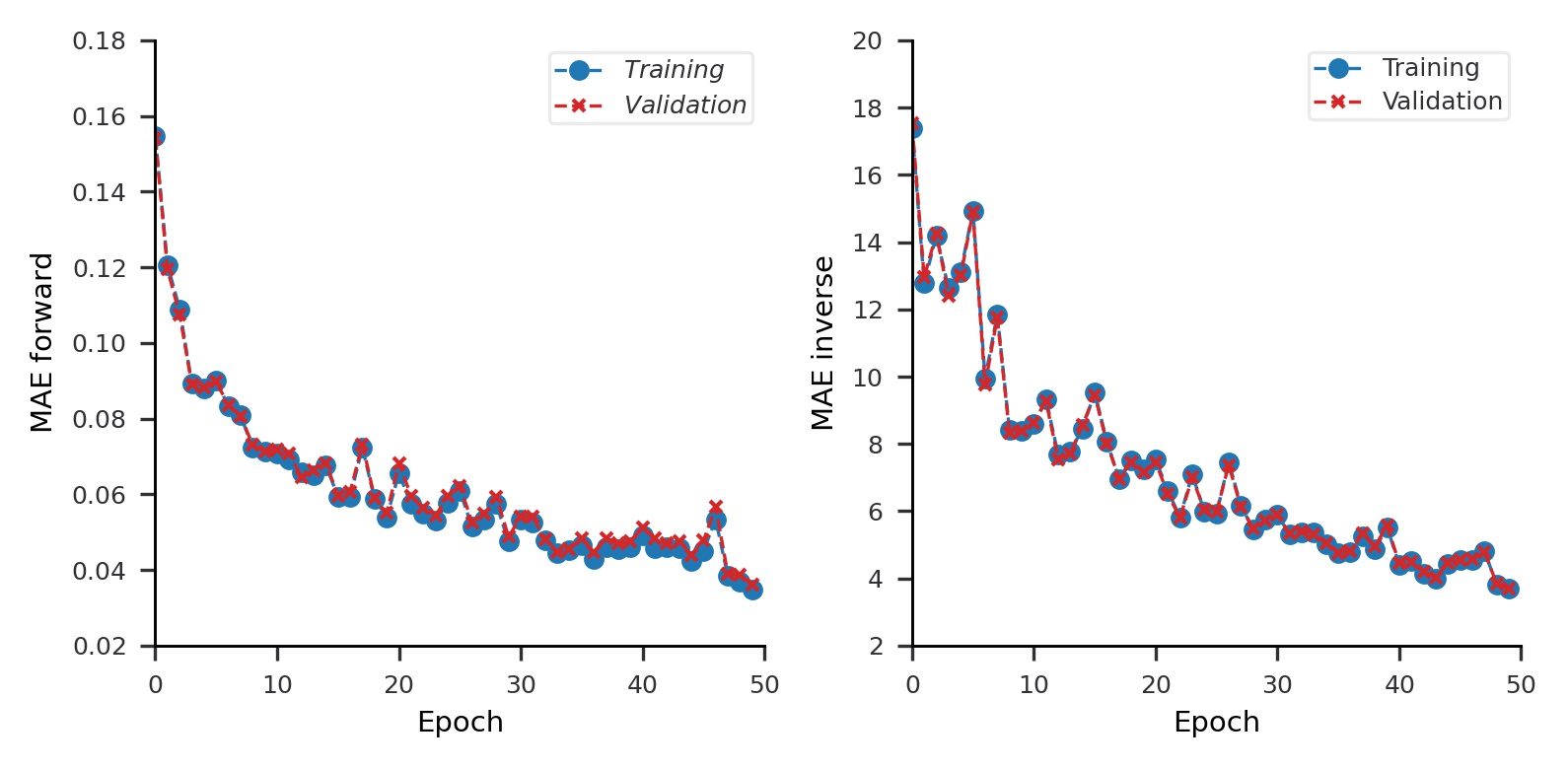}
    \caption{Performance of the best model in forward and inverse direction in terms of mean absolute error (MAE) over the number of epochs. The blue dots denote the MAE for the training data and the red crosses for the validation data.}
    \label{fig:history_uni}
\end{figure}

The first case, that is considered, is the monomodal case with all available training and validation data used, i.e. three wavelengths, $P_{11}$ and $-\frac{P_{12}}{P_{11}}$ and all angles. The $R^2$ value for the validation data set for the inverse process at the last epoch of training is $R^2= 0.993$. The $R^2$ value for the forward pass scored on the validation data set is $R^2 = 0.9978$. So, for both directions, the values are very close to the optimal and theoretical maximum value of $1$. In Figure \ref{fig:history_uni} the history of the mean absolute error (MAE) for the forward (left) and inverse (right) model over all epochs is depicted. The tendency of the curves is clearly decreasing, this means, that the models in forward and inverse direction are getting better the more epochs are used. In addition, the blue (training data) and red (validation) curves overlap until 50 epochs, so there is no sign of overfitting visible in this plot. 

This best model is then used for testing with the so far unseen data (i.e., with the test data set consisting of 20 000 data points). For the forward model this results in a maximum relative error for $\log(P_{11})$ of $1.5\%$ and a maximum absolute error of $0.06$ for $-\frac{P_{12}}{P_{11}}$ both at confidence interval of $95\%$. The mean $\operatorname{wMAPE}$s are given by $\operatorname{wMAPE}(P_{11})= 4.29\%$ and $\operatorname{wMAPE}(-\frac{P_{12}}{P_{11}}) = 4.20\%$. These results compare very favourably with respect to possible real measurement device errors, which we take as $5\%$ relative error for $P_{11}$ and $0.1$ absolute error for $-\frac{P_{12}}{P_{11}}$ as explained in Section \ref{space_variable_y}.

%\begin{table}[]
%\begin{tabular}{lrrrrrrr}
%Aerosol Properties& $V_{tot}$ & $R_{mean}$& $\sigma_g$ & n & $k_{450}$ &$k_{532}$& $k_{630}$ \\
%Mean Abs. Error & 22.04 & 0.0067 & 0.0015 & 0.0013 & 0.0013 & 0.00078 & 0.00060 \\
%Std. Abs. Error & 20.25 & 0.0060 & 0.0018 & 0.0011 & 0.0013 & 0.00097 & 0.00099 \\
%Abs. Error 95\% max. & 58.55 & 0.02 & 0.005 & 0.003 & 0.004 & 0.002 & 0.002\\
%$\operatorname{wMAPE}$[\%] & 0.88 & 0.50 & 0.11 & 0.09 & 1.34 & 1.23 & 1.40
%\end{tabular}
%\caption{Results for the aerosol property retrieval for the unimodal data set with three wavelengths and $P_{11}$ and $-\frac{P_{12}}{P_{11}}$ and all angles.}
%\label{tab:inv_pass}
%\end{table}

\begin{table}[]
\begin{tabular}{lrrrrrrr}
Aerosol Properties& $V_{tot}$ & $R_{mean}$& $\sigma_g$ & n & $k_{450}$ &$k_{532}$& $k_{630}$ \\
Mean Abs. Error & 22.04 & 0.0067 & 0.0015 & 0.0013 & 0.0013 & 0.00078 & 0.00060  \\
Abs. Error 95\% max. & 58.55 & 0.02 & 0.005 & 0.003 & 0.004 & 0.002 & 0.002\\
$\operatorname{wMAPE}$[\%] & 0.88 & 0.50 & 0.11 & 0.09 & 1.34 & 1.23 & 1.40 \\
$\operatorname{wMAPE}$[\%] noise & 1.58 & 1.01 & 0.34 & 1.93 & 2.15 & 1.93 & 2.17
\end{tabular}
\caption{Aerosol property retrieval error results in terms of the Mean Absolute Error, the maximal Absolute Error at a confidence level of 95\%, the $\operatorname{wMAPE}$ and the $\operatorname{wMAPE}$ for the case of noisy test data (assuming 5\% Gaussian noise), for the monomodal data set with ideal measurement configuration ($P_{11}$ and $-\frac{P_{12}}{P_{11}}$ at three wavelengths and all angles).}
\label{tab:inv_pass}
\end{table}

For the inverse model, the results of the aerosol property retrieval are summarized in Table \ref{tab:inv_pass}. There the mean absolute error and the maximal absolute error at a confidence level of 95\% as well as the $\operatorname{wMAPE}$ are stated. The $\operatorname{wMAPE}$ for all aerosol properties stays below 1.5\%. The aerosol properties that are predicted best, according to the $\operatorname{wMAPE}$s, are the real part of the refractive indices and the geometric standard deviation,  whereas the parameters that are predicted worst, are the complex parts of the refractive indices. Also in the case of noisy test data, here we assumed 5\% Gaussian noise added to the test data, the $\operatorname{wMAPE}$ is stated in Table \ref{tab:inv_pass}. It stays below 2.2\% for all aerosol properties, which clearly shows that the proposed method for aerosol property retrieval is robust against possible random measurement noise. 

\begin{figure}
    \centering
    \includegraphics[width=\textwidth]{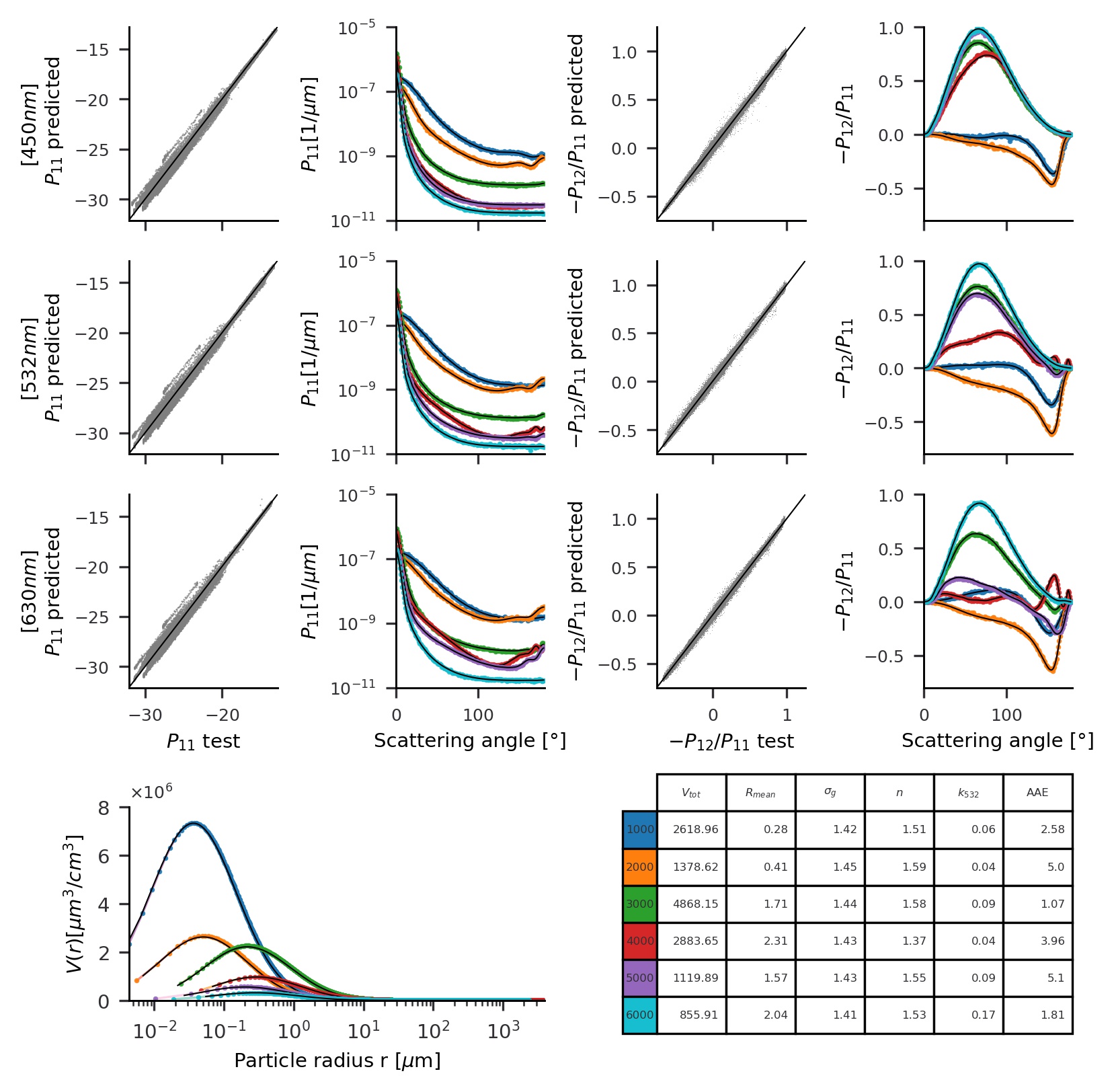}
    \caption{Comparison of the predicted and simulated test data. First and third column show correlation plots of the predicted versus the test data values of the (polarized) phase functions for all three wavelengths. Second and fourth column depict the simulated (black line) versus predicted (polarized) phase functions (colored dots) in a qualitative representation. Six different data samples were chosen. In the last line, the qualitative representation of the simulated versus the predicted total volume concentration over the particle radii for the according six samples is shown. The table in the lower right corner states the simulated aerosol property values of these six samples. }
    \label{fig:test_pred_uni}
\end{figure}

The qualitative comparison of the forward prediction for different wavelengths and the inverse prediction of the particle size distribution and the according test data for six randomly chosen data samples each are depicted in Figure \ref{fig:test_pred_uni}. Although the forward prediction follows the path of the test data quite good, they look a bit noisy, as can be seen in the figure (colored curves), nevertheless, this is due to the nature of the used prediction method, i.e. each data point of the simulated data set is seen as a single point and not part of a mathematical function, which would make the curve looking smoother. The colors across all images in this figure belong together and refer to the aerosol property values stated in the table in the lower right edge of the figure. In addition, the correlation plots for the whole test data set are shown for the predicted versus the test data for the (polarized) phase function at different wavelengths. In general, a good fit for all the data can be seen here, since all the grey dots are located close to the black line, which marks a perfect fit between test data and prediction.   

Concerning the computation time, one forward prediction with the INN model takes approximately 0.46 ms, compared to GRASP-OPEN which took around 4 s for these simulations. For one inverse prediction, the INN model needs around 2.64 ms, compared to approximately 6 s for the GRASP-OPEN inversion. 
So for the monomodal case with all possible measurement data available, the proposed invertible neural network turns out to be fast and accurate.

\subsection{Monomodal case with imperfect measurement configurations: results with missing angles, wavelengths and polarimetric information}

In this section, the results from the case studies inspired by real measurements are presented for the monomodal aerosol data set. As stated in section \ref{space_variable_y} the polar nephelometer usually can not measure all angles, hence in these case studies, all neural networks are trained assuming that the angles 0°-5°, 85°-95°, 175°-179° are missing in the data set.
Furthermore, it can happen that not both the phase functions and polarized phase functions can be measured at all 3 wavelengths. Hence, we compare networks trained with either $P_{11}$ or $-\frac{P_{12}}{P_{11}}$, at either 1, 2 or 3 wavelengths to see the performance of the neural network under different measurement data conditions (see Table \ref{tab:case_study}). 

\begin{figure}
    \centering
    \includegraphics[width=1\textwidth]{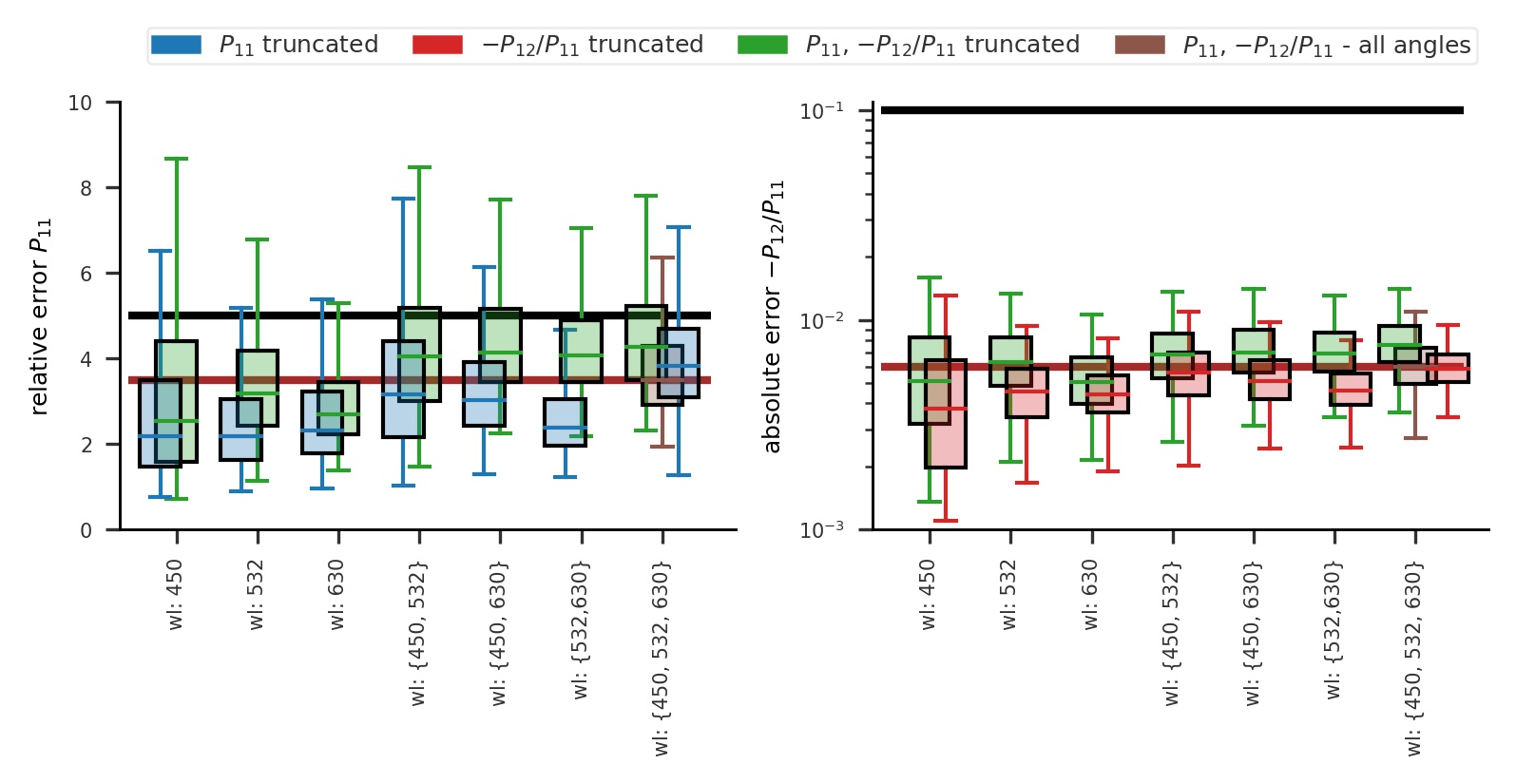}
    \caption{Forward model performance: Box plots of the relative error for the phase function and the absolute errors for the polarized phase functions for all 22 models for the monomodal case. Each model is depicted by a box extending from the first quartile to the third quartile and a horizontal line running through the box  at the median. The whiskers show the range of the absolute errors over the 20 000 data points in the test set. The brown solid lines indicate the median errors of the full data case, whereas the solid black lines show the assumed measurement device errors.}
    
    \label{fig:qoi_abs_err_uni}
\end{figure}

In forward direction the performance of all models is similar, the $R^2$ values for the test data are at least $0.997$ as can be seen in Figure \ref{fig:r2_test_qoi} in the \ref{App:DetailedResults}. This is also visible in the box plots of the errors of the forward models shown in Figure \ref{fig:qoi_abs_err_uni}. On the x-axis the different wavelength combinations are depicted and on the y-axis the relative errors for $P_{11}$ and the absolute errors for $-\frac{P_{12}}{P_{11}}$. The colors denote whether the INN was trained to predict only truncated $P_{11}$ (blue), only truncated $-\frac{P_{12}}{P_{11}}$ (red), both truncated phase functions (green) or both phase functions without angle truncation (brown). The errors are lower, when only single type functions (red and blue) are used, compared to the models that use all the available functions. This is most likely because, for the case with two functions available, the amount of data points is doubled. Hence, not only more failures can happen, but probably also more training data or an adaption of the INN layers would be necessary to get the same performance for all cases.

The solid brown lines indicate the median errors of the full data case and one sees that all the models behave similarly. To place the retrieval errors in context the figure also displays target error limits (solid black lines), which are defined as the corresponding measurement errors expected in $P_{11}$ and $-\frac{P_{12}}{P_{11}}$ (i.e., 5\% and 0.1, respectively). It is clear that for all wavelength combinations the INN forward model retrieval errors are within the measurement error limits, which indicates that the error associated with the INN models are minor.

\begin{figure}
    \centering
    \includegraphics[width=\textwidth]{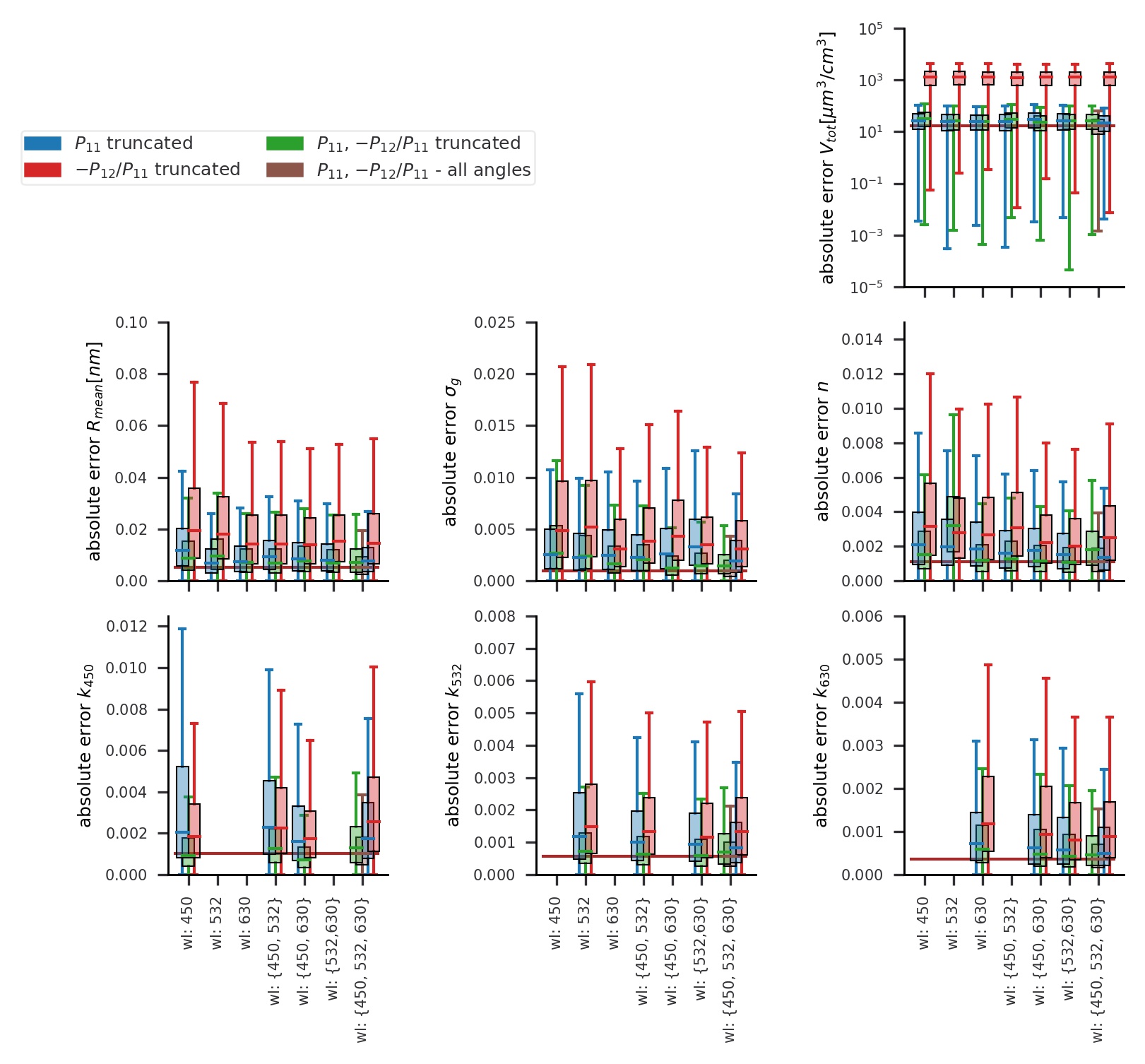}
    \caption{Inverse model performance: Boxplots of the absolute errors of each aerosol property. Each box describes the lower and upper quartile values and the median(line). The range of the absolute errors over all data points in the test set is shown with the whiskers.}
    \label{fig:abs_uni}
\end{figure}

In Figure \ref{fig:abs_uni} the boxplots of the mean absolute errors for the aerosol properties, so the inverse prediction are depicted for comparison among the different cases. Each subplot displays the mean absolute errors for another aerosol quantity. It can be seen that the volume concentration can't be predicted at all if only the normalized polarized phase functions (red) are available. This is expected since absolute light intensity information (which is strongly proportional to the aerosol volume concentration) is lost in the ratio of $-\frac{P_{12}}{P_{11}}$. Also the retrieval of the radius $R_{mean}$, the real part of the refractive index $n$ and the geometric standard deviation $\sigma_g$ is worse, if only $-\frac{P_{12}}{P_{11}}$ is at hand. Whereas for the complex part of the refractive index it seems, that both functions are equally suited for prediction. In general, for all aerosol properties the best results are gained if both functions are accessible. Concerning the wavelengths, the absolute error is the biggest, if only one wavelength is used. Whereas the retrieval works best if all three wavelengths can be utilized. For the three wavelengths case with $P_{11}$ and $-\frac{P_{12}}{P_{11}}$, the network performs just slightly better if all angles can be measured. In general, one sees that increasing the measurement space is an advantage in the inverse model, but not in the forward model. Summarized, it can be emphasized, that except the case with only $-\frac{P_{12}}{P_{11}}$, all other cases give similar results as the full data case and can hence be used also in practice for the retrieval of aerosol properties for the application involving a monomodal homogeneous aerosol. 


\subsection{Bimodal case with ideal measurement configuration: $P_{11}$, $-\frac{P_{12}}{P_{11}}$ at all 3 wavelengths and all angles}

In this section, the results for the more complex bimodal aerosol test case, see Table \ref{tab:bimodal_variables}, with all available data are presented (i.e., assuming an ideal measurement configuration). The results are similar to the more simple monomodal case (see Section \ref{sec:results-monomodal-ideal}). 
The coefficient of determination of the validation data set in forward direction is $R^2 = 0.9975$ and for the inverse pass $R^2= 0.985$.
\begin{figure}
    \centering
    \includegraphics[width=\textwidth]{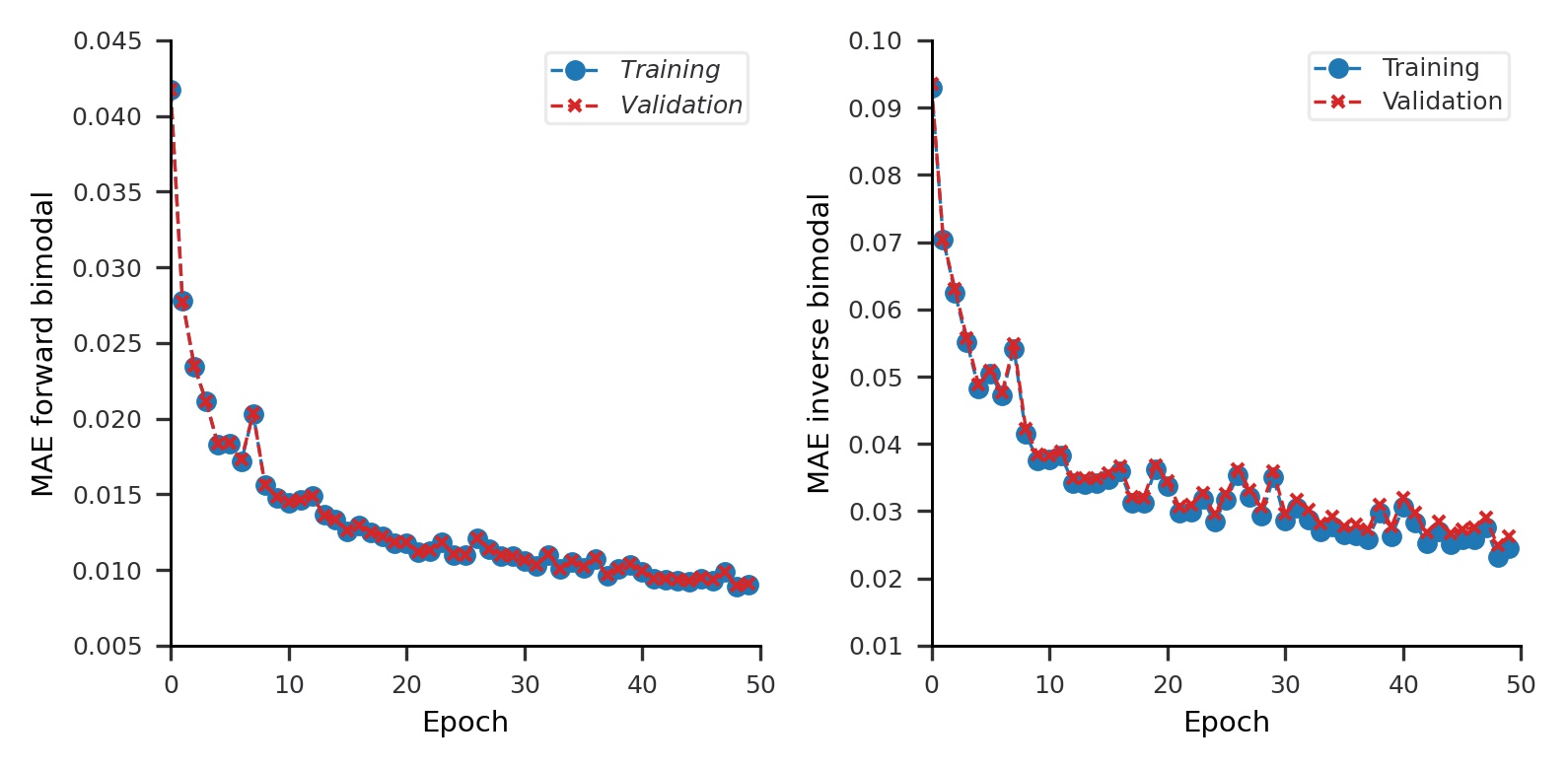}
    \caption{Performance of the best model in forward and inverse direction in terms of mean absolute error (MAE) over the number of epochs for the bimodal case. The blue dots denote the MAE for the training data and the red crosses for the validation data.}
    \label{fig:history_bim}
\end{figure}
Figure \ref{fig:history_bim} shows the mean absolute error for the best inverse and forward model for training and validation data evolving along the epochs, which is in both cases clearly decreasing, meaning that the model improves over the number of epochs. This best trained model is validated with the so far unseen test data set. The maximum relative error for $log(P_{11})$ is $0.33\%$ and the maximum absolute error for $-\frac{P_{12}}{P_{11}}$ is $0.03$ at a confidence level of 95\%. The $\operatorname{wMAPE}$s are as following: $\operatorname{wMAPE}(P_{11})=1.34\%$ and $\operatorname{wMAPE}(-\frac{P_{12}}{P_{11}})=1.89\%$. Also in the bimodal case, the prediction errors are lower than the assumed device measurement errors and are even better than for the monomodal case. The mean absolute errors and the maximal absolute errors at a 95\% confidence level and the $\operatorname{wMAPE}$s for all aerosol properties are stated in Table \ref{tab:inv_pass_bim}. The properties that are retrieved best according to the $\operatorname{wMAPE}$ are the real parts of the refractive indices followed by the geometric standard deviations and the radii for both, the coarse and fine mode. Again the complex parts of the refractive indices for all wavelengths and all modes are predicted worst. But in general, the $\operatorname{wMAPE}$ stays below 3.4\% for all quantities.  This shows that, first, the spectral polarized light scattering phase functions also contain ample information on aerosol absorptive properties and, second, that the INN is capable of retrieving these. However, in practical applications the retrieval of aerosol absorptive properties from phase function measurements will often be hampered by the fact that a simplified representation of the state space is typically inappropriate in the presence of light absorbing black carbon particles, see e.g. \cite{Schuster_2019}.
For completeness, the correlation plots of test and predicted test data for the forward model and the qualitative comparison between test and predicted test data for the forward model and the bimodal particle size distribution for six randomly chosen test data is depicted in Figure \ref{fig:test_pred_bim} in the \ref{App:DetailedResults}. The according values of the particle properties are given  in Table \ref{fig:pred_plot_values_bim} in the \ref{App:DetailedResults}.

%\begin{table}[]
%\begin{tabular}{lrrrrrr}
%Aerosol Properties& $V_{tot}$ & $R_{mean_{fine}}$ & $R_{mean_{coarse}}$ & $\sigma_{g_{fine}}$ & $\sigma_{g_{coarse}}$ & $\chi$\\
%Mean Abs. Err. & 0.12 & 0.00018 & 0.0031 & 0.0018 & 0.0021 & 0.0023 \\
%Mean Abs. Err. stdv. & 0.099 & 0.00018 & 0.0031 & 0.0018 & 0.0012 & 0.0019 \\
%Abs. Err. 95\% max. & 0.305 & 0.0005 & 0.0084 & 0.0046 &0.0056 & 0.006 \\
%$\operatorname{wMAPE}$ [\%] & 0.694 & 0.127 & 0.283 & 0.124 & 0.147 & 0.319 \\
%\hline\hline
%Aerosol Properties &$k_{450_{fine}}$ &$k_{450_{coarse}}$ &$k_{532_{fine}}$&$k_{532_{coarse}}$ & $k_{630_{fine}}$&$k_{450_{coarse}}$  \\
%Mean Abs. Err. & 0.0020 & 0.0003 & 0.0013 & 0.00026 & 0.0012 & 0.00024 \\
%Mean Abs. Err. stdv. & 0.0020 & 0.00031 & 0.0012 & 0.00027 & 0.0014 & 0.00025 \\ 
%Abs. Err. 95\% max. & 0.0053 & 0.0009 & 0.0037 & 0.0007 & 0.0037 & 0.0007\\
%$\operatorname{wMAPE}$ [\%] & 1.988 & 3.003 & 2.118 & 3.087 & 2.745 & 3.336\\
%\hline\hline
%Aerosol Properties & $n_{fine}$ & $n_{coarse}$ & $\phi$ & $AAE_{fine}$ & $AAE_{coarse}$ & \\
%Mean Abs. Err. & 0.0018 & 0.0012 & 0.0057 & 0.14 & 0.07 \\
%Mean Abs. Err. stdv. & 0.0014 & 0.0014 & 0.0051 & 0.27 & 0.086 \\
%Abs. Err. 95\% max. & 0.0044 & 0.0035 & 0.0151 & 0.480 & 0.237 & \\
%$\operatorname{wMAPE}$ [\%] & 0.124 & 0.08 & 1.109 & 3.369 & 3.48 &
%\end{tabular}
%\caption{Results for the aerosol property retrieval for the bimodal data set with three wavelengths, all angles and $P_{11}$ and $-\frac{P_{12}}{P_{11}}$.}
%\label{tab:inv_pass_bim}
%\end{table}

\begin{table}[]
\begin{tabular}{lrrrrrr}
Aerosol Properties& $V_{tot}$ & $R_{mean_{fine}}$ & $R_{mean_{coarse}}$ & $\sigma_{g_{fine}}$ & $\sigma_{g_{coarse}}$ \\
Mean Abs. Err. & 0.12 & 0.00018 & 0.0031 & 0.0018 & 0.0021  \\
Abs. Err. 95\% max. & 0.305 & 0.0005 & 0.0084 & 0.0046 &0.0056  \\
$\operatorname{wMAPE}$ [\%] & 0.694 & 0.127 & 0.283 & 0.124 & 0.147  \\
$\operatorname{wMAPE}$ [\%] noisy & 2.24 & 0.62 & 0.98 & 0.46 & 0.72 \\
\hline\hline
Aerosol Properties &$k_{450_{fine}}$ &$k_{450_{coarse}}$ &$k_{532_{fine}}$&$k_{532_{coarse}}$ & $k_{630_{fine}}$  \\
Mean Abs. Err. & 0.0020 & 0.0003 & 0.0013 & 0.00026 & 0.0012 \\
Abs. Err. 95\% max. & 0.0053 & 0.0009 & 0.0037 & 0.0007 & 0.0037 \\
$\operatorname{wMAPE}$ [\%] & 1.988 & 3.003 & 2.118 & 3.087 & 2.745 \\
$\operatorname{wMAPE}$ [\%] noisy & 6.47 & 8.67 & 7.21 & 8.56 & 8.87 \\
\hline\hline
Aerosol Properties &$k_{630_{coarse}}$& $n_{fine}$ & $n_{coarse}$ & $\phi$ & $\chi$  \\
Mean Abs. Err.& 0.00024  & 0.0018 & 0.0012 & 0.0057 & 0.0023 \\
Abs. Err. 95\% max. & 0.0007& 0.0044 & 0.0035 & 0.0151 & 0.006  \\
$\operatorname{wMAPE}$ [\%] & 3.336& 0.124 & 0.08 & 1.109 & 0.319 \\
$\operatorname{wMAPE}$ [\%] noisy & 8.77 & 0.41 & 0.28 & 3.58 & 1.34
\end{tabular}
\caption{Results in terms of the Mean Absolute Error, the maximal Absolute Error at a confidence level of 95\%, the $\operatorname{wMAPE}$ and the $\operatorname{wMAPE}$ for the case of noisy test data, assuming 5\% Gaussian noise,  for the aerosol property retrieval for the bimodal data set with three wavelengths, all angles and $P_{11}$ and $-\frac{P_{12}}{P_{11}}$.}
\label{tab:inv_pass_bim}
\end{table}

Table \ref{tab:inv_pass_bim} also shows the $\operatorname{wMAPE}$, when 5\% Gaussian noise is added to the test data, which is closer to real measurements, than noise-free test data ($\operatorname{wMAPE}$ [\%] noisy). The $\operatorname{wMAPE}$ for all aerosol properties stays below 9\% and still the best predicted properties are the radii, the geometric standard deviations and the real parts of the refractive indices for both modes, with $\operatorname{wMAPE}$ below 1\%. This shows that the architecture of the invertible neural network is robust against random measurement errors. 

The computation times for one forward and inverse pass are similar to the monomodal case, 0.46 ms and 2.66 ms, respectively. 
Hence, as in the monomodal case, also in the bimodal case, the invertible neural network method proofs to be an accurate, robust and fast retrieval method for aerosols from measurement data.

\subsection{Bimodal case with imperfect measurement configurations: results with missing angles, wavelengths and polarimetric information}

For the bimodal case, again the same case study is performed as in the monomodal case, meaning that networks were trained, with different numbers and values of wavelengths, different presence of the (polarized) phase functions and missing angles (0°-5°, 85°-95°, 175°-179°). For the forward pass the comparison of the $R^2$ values shows as expected, that the prediction is similarly good for all models, see Figure \ref{fig:r2_test_qoi_bim} in the \ref{App:DetailedResults}. The boxplots of the relative and mean absolute errors for the forward model are depicted in Figure \ref{fig:abs_err_qoi_bim} as well in the Appendix. The errors show a similar behaviour over the different cases as for the monomodal data set. Again the results for all models are much better compared to the assumed measurement device errors (solid black lines).

\begin{figure}
    \centering
    \includegraphics[width=\textwidth]{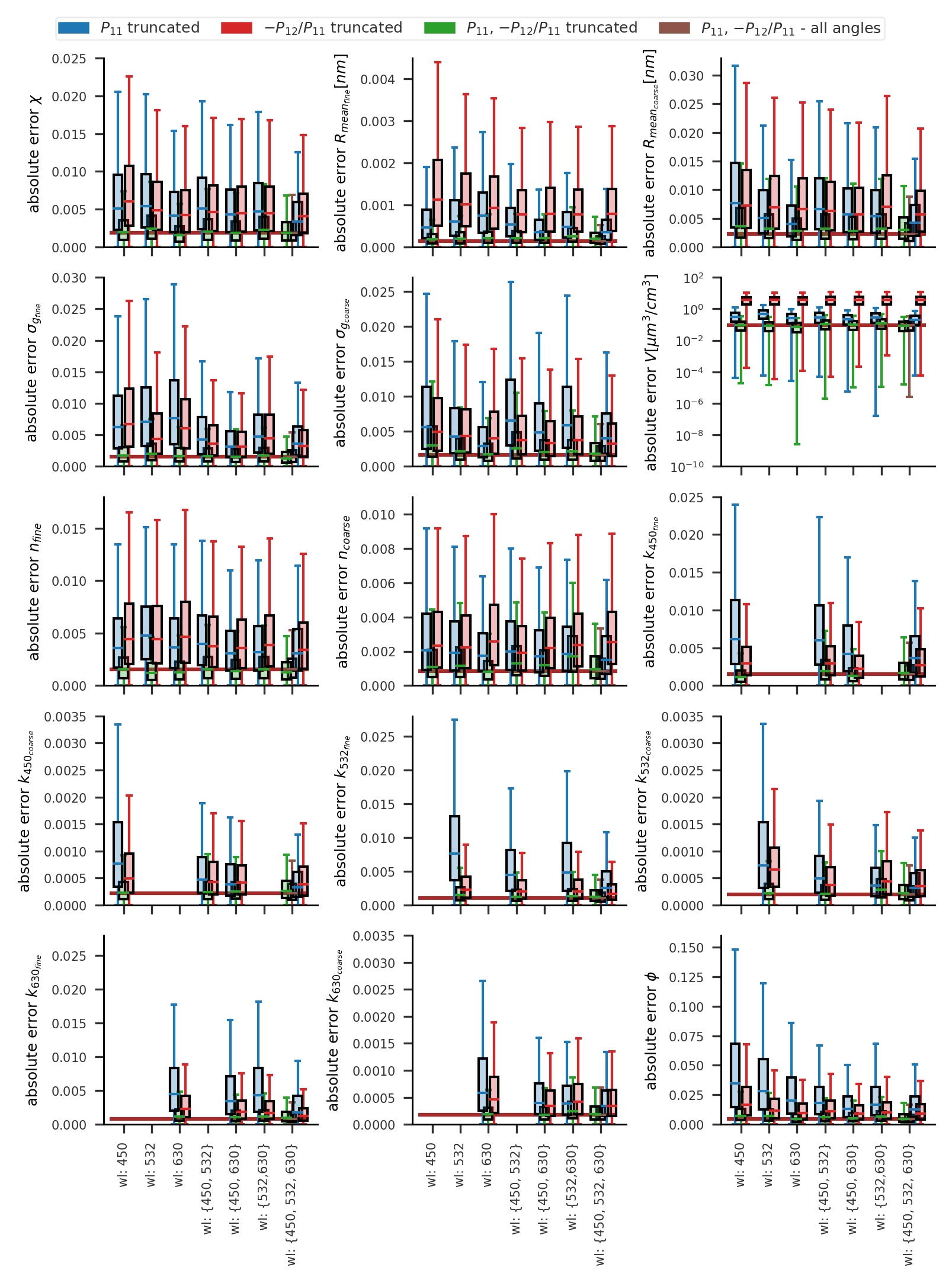}
    \caption{Boxplots of the absolute errors of each aerosol property for the bimodal case. Each box describes the lower and upper quartile values and the median(line). The range of the absolute errors over all data points in the test set is shown with the whiskers.}
    \label{fig:abs_bim}
\end{figure}
For the inverse pass, as before, it can be clearly seen at the boxplots in Figure \ref{fig:abs_bim} of the absolute errors, that solely the polarized phase function is not suited for the prediction of the volume concentration, independently of the wavelength, for the reason already explained for the monomodal examples. In general the absolute error is smallest for all aerosol properties, if all 3 wavelengths, $P_{11}$ and $-\frac{P_{12}}{P_{11}}$ are available. The missing angles hardly influence the retrieval. 
Even if not all the wavelengths can be measured, it is important to have the polarized phase functions and the phase functions available, because the mixture of these two decreases the retrieval uncertainty for all aerosol properties. For the prediction of the radius, $R_{mean_{fine}}$, and the real part of the refractive index, $n_{fine}$, $P_{11}$ seems to be better suited, whereas $-\frac{P_{12}}{P_{11}}$ solely, gives better results for the spherical fraction and all the complex parts of the refractive indices. No significant statement can be drawn, whether one or the other wavelength, or a combination of two wavelengths is better for the aerosol retrieval.
Also in this bimodal case study it turned out that except solely $-\frac{P_{12}}{P_{11}}$ all other combinations of possible data are suited for the retrieval of aerosol properties from measurement data.

\section{Conclusion, Discussion and Outlook}
In summary, we introduced a novel method for aerosol property retrieval from in situ measurement data using invertible neural networks. The special structure of the neural networks allows to not only retrieve the aerosol properties, so solve the inverse problem, but also to simulate the forward model, which is to calculate measurement data from aerosol properties. By simulating laboratory and field in situ measurements with GRASP-OPEN and by using them for training and testing, we have shown the practical applicability of the proposed method. The quality achievable with the forward model is sufficient, since the current measurement device errors exceed the errors introduced by the model. This demonstrates that when it comes to application to atmospheric aerosols with complex size distribution and mixing state, the major errors will arise from the need to choose a simplified state space to represent the aerosol (i.e., from the ill-posedness of the inverse problem) rather than from limitations of INN capability. In addition, one forward simulation of the INN lasts not only a millisecond, which is much faster compared to physics based simulations that typically require seconds for one simulation. Also the performance of the inverse model, so the retrieval of the aerosol properties turned out to be satisfying, with a weighted mean absolute percentage error for all aerosol properties except the complex part of the refractive index from monomodal and bimodal case staying below 1.2\% (and for the complex part of the refractive index below 3.4\%), and a simulation time for one retrieval 1000 times faster than e.g. with GRASP-OPEN. This enables the near-real-time usage of the aerosol property retrieval from measurements and hence, could be a step further for processing of data from new sensors in real-time.

To reproduce real measurement conditions, we tested the method with data including 5\% Gaussian measurement noise. The results showed that the method is robust against measurement errors, meaning that the errors for the retrieval in terms of the weighted mean absolute percentage error still stay below 3.6\% for all aerosol properties except the complex part of the refractive index, for which it is less than 9\%. 
In addition, a case study was performed to see how the models deal with missing data. Therefore, neural networks were trained and tested for 22 different cases of missing angles, wavelengths and/or (polarized) phase functions. Although the results for the retrieval are best if all data are available, nearly all the models with missing data achieved comparably good results, such that they are useful in practice. Additional noteworthy results include the fact that the addition of polarized phase function information offers a distinct performance benefit for retrieval of particle morphology information (i.e., the fraction of non-spherical particles in the coarse mode).

This proves, that invertible neural networks are a fast, accurate and robust method to retrieve aerosol properties from in situ, multi-angle light scattering measurements. Hence, invertible neural networks seem to be a promising alternative to commonly used pre-computed look-up tables and iterative, physics-based inversion methods. 

After all, we see some options to improve the methods and datasets for future work:
If a need for higher accuracy arose, the model architecture itself could be improved, by allowing different neural networks in the affine coupling blocks or by performing a more sophisticated hyperparameter scan.
If physical knowledge is at hand, it is advisable to incorporate this, e.g. adding barrier functions to the loss to guarantee that physical quantities are within reasonable physical ranges or adding loss terms including Mie scattering theory, so using physics informed neural networks, \cite{RAISSI2019686}.
Additionally, in further studies we aim to relax some of the simplifying assumptions concerning the data set that we have used here. Like for example allowing multiple state parameters describing $n(\lambda)$ in order to capture any spectral dependence, even if such dependence is only minor. Given the good results presented here and the fact that iterative models such as GRASP-OPEN are capable of capturing such spectral dependence, we expect that INN models will also be capable of capturing this.
As discussed above, we considered here a relatively simplified bimodal representation of atmospheric aerosol size distributions where the fine and coarse aerosol modes are well-separated, which is consistent with ambient measurements processed with a state-of-the-art in classical retrieval scheme (GRASP-OPEN) \cite{Espinosa_2019}. In future work, we intend to explore inversion performance also for the cases of overlapping size distribution modes, as well as for the case when the size distribution is represented in a sectional manner with a greater number of state parameters.

\section{Funding source}

Romana Boiger was funded by the PSI Career Return Program. Financial support was also received from MeteoSwiss through a science project in the framework of the Swiss contribution to the global atmosphere watch programme (GAW-CH).

\section{Acknowledgement}
We acknowledge
Oleg Dubovik, Tatsiana Lapionak, Anton Lopatin and David Fuertes (GRASP-SAS, Remote sensing developments, Université de Lille, 59655 Villeneuve d'Ascq Cedex, France) for their support with GRASP-OPEN.
%% The Appendices part is started with the command \appendix;
%% appendix sections are then done as normal sections
\appendix

\section{Results: Definitions and Details}
%\label{sec:sample:appendix}
\subsection{Preprocessors}
Let $X$ be a set of data. Then the following scaler can be applied to get a scaled version of this data set $X_{scaled}$:
\begin{itemize}
\item \textbf{Adaptive Min Max Scaler}
\begin{align*}
    X_{scaled} &= X_{std}(X_{max}-X_{min})+ X_{min} \\
    X_{std} &= \frac{X-X_{min}}{X_{max}-X_{min}}
\end{align*}
\item \textbf{Standard Scaler}
\begin{align*}
X_{scaled} = \frac{X-X_{mean}}{X_{std}}
\end{align*}
\end{itemize}
\subsection{Metrics} \label{App:Metrics}
Let $y=(y_1, ... , y_n)$ be the true value and $\hat{y}={\hat{y}_1,...,\hat{y}_n}$ be the predicted value. 
\begin{itemize}
    \item Coefficient of Determination $R^2(y,\hat{y}) = 1-\frac{\sum^n_{i=1}(y_i - \hat{y}_i)^2}{\sum^n_{i=1}(y_i-\bar{y})^2}$, where $\bar{y}=\frac{1}{n}\sum^{n}_{i=1}y_i$. The best possible score is 1.
    \item Weighted Mean Absolute Percentage Error $\operatorname{wMAPE} = \frac{\sum^n_{i=1}|y_i-\hat{y}_i|}{\sum^n_{i=1}|y_i |}$ . The advantage of $\operatorname{wMAPE}$ compared to $\operatorname{MAPE}$, which is used quite often, is, that division by zero or a value close to zero is avoided, by taking the sum over all actual values instead of taking just one actual value in the denominator. 
    \item Absolute Error: $AE =|y-\hat{y}|$
    \item Relative Error: $RE =\frac{ |y-\hat{y}|}{y}$
\end{itemize}

\subsection{Case Study}
Table \ref{tab:case_study} lists all the cases used in the case studies in the results section. 

\begin{table}[]
\begin{tabular}{lrrrrrr}
Cases & $P_{11}$ & $-P_{12}/P_{11}$ & wl 450 nm & wl 532 nm & wl 630 nm & all angles \\
1 & x & x & x & x & x & x \\
2 & x & x & x & x & x &  \\
3 & x & x & x &  &  &  \\
4 & x & x &  & x &  &  \\
5 & x & x &  &  & x &  \\
6 & x & x & x &  & x &  \\
7 & x & x & x & x &  &  \\
8 & x & x &  & x & x &  \\
9 & x &  & x &  &  &  \\
10 & x &  &  & x &  &  \\
11 & x &  &  &  & x &  \\
12 & x &  & x &  & x &  \\
13 & x &  & x & x &  &  \\
14 & x &  &  & x & x &  \\
15 & x &  & x & x & x &  \\
16 &  & x & x &  &  &  \\
17 &  & x &  & x &  &  \\
18 &  & x &  &  & x &  \\
19 &  & x & x &  & x &  \\
20 &  & x & x & x &  &  \\
21 &  & x &  & x & x &  \\
22 &  & x & x & x & x &  \\
\end{tabular}
\caption{Case study}
\label{tab:case_study}
\end{table}

\subsection{Detailed Results}\label{App:DetailedResults}

\begin{figure}
    \centering
    \includegraphics[width=1\textwidth]{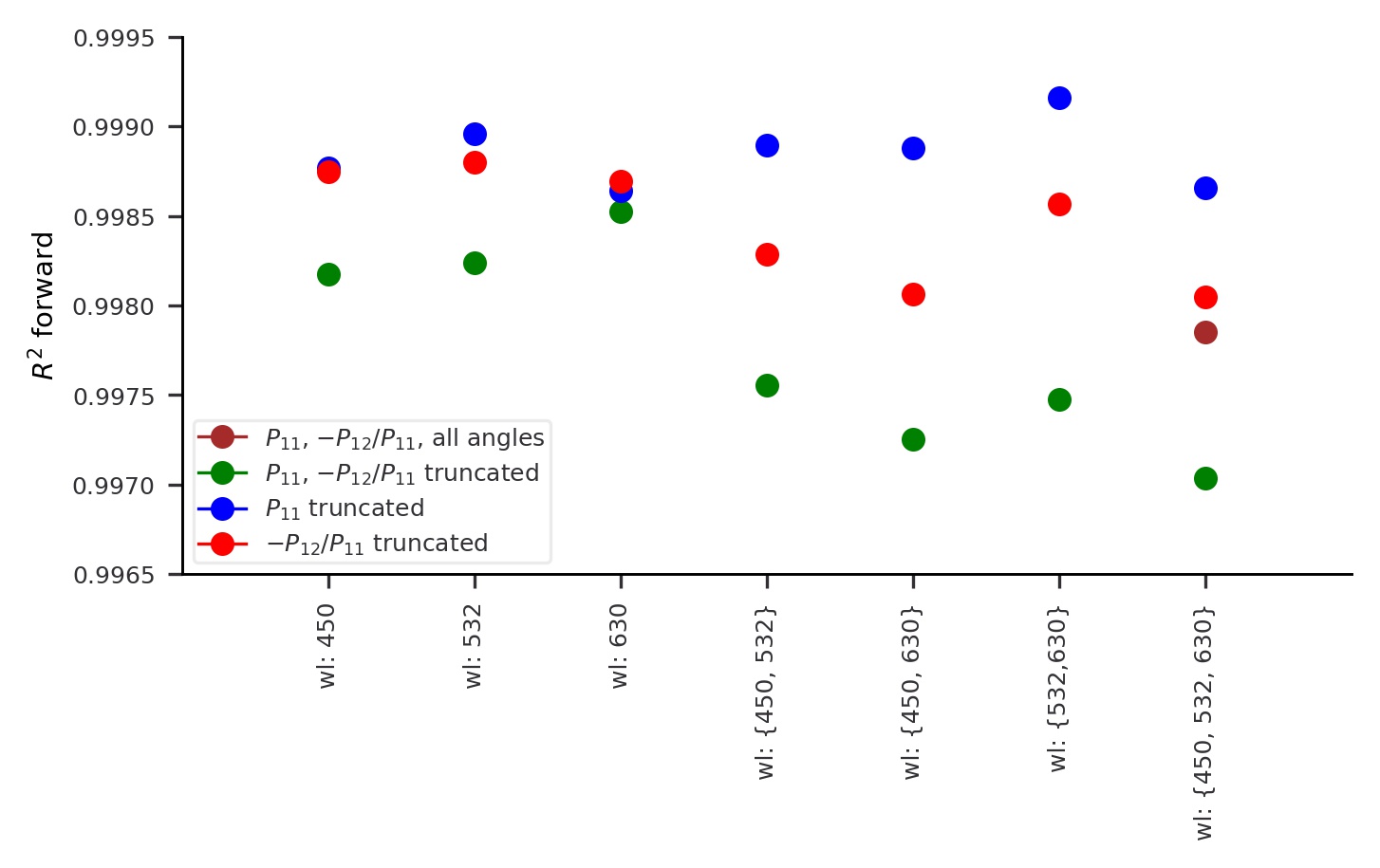}
    \caption{The $R^2$ values for the forward prediction for the different cases for the monomodal data set.  }
    
    \label{fig:r2_test_qoi}
\end{figure}

\begin{figure}
    \centering
    \includegraphics[width=1\textwidth]{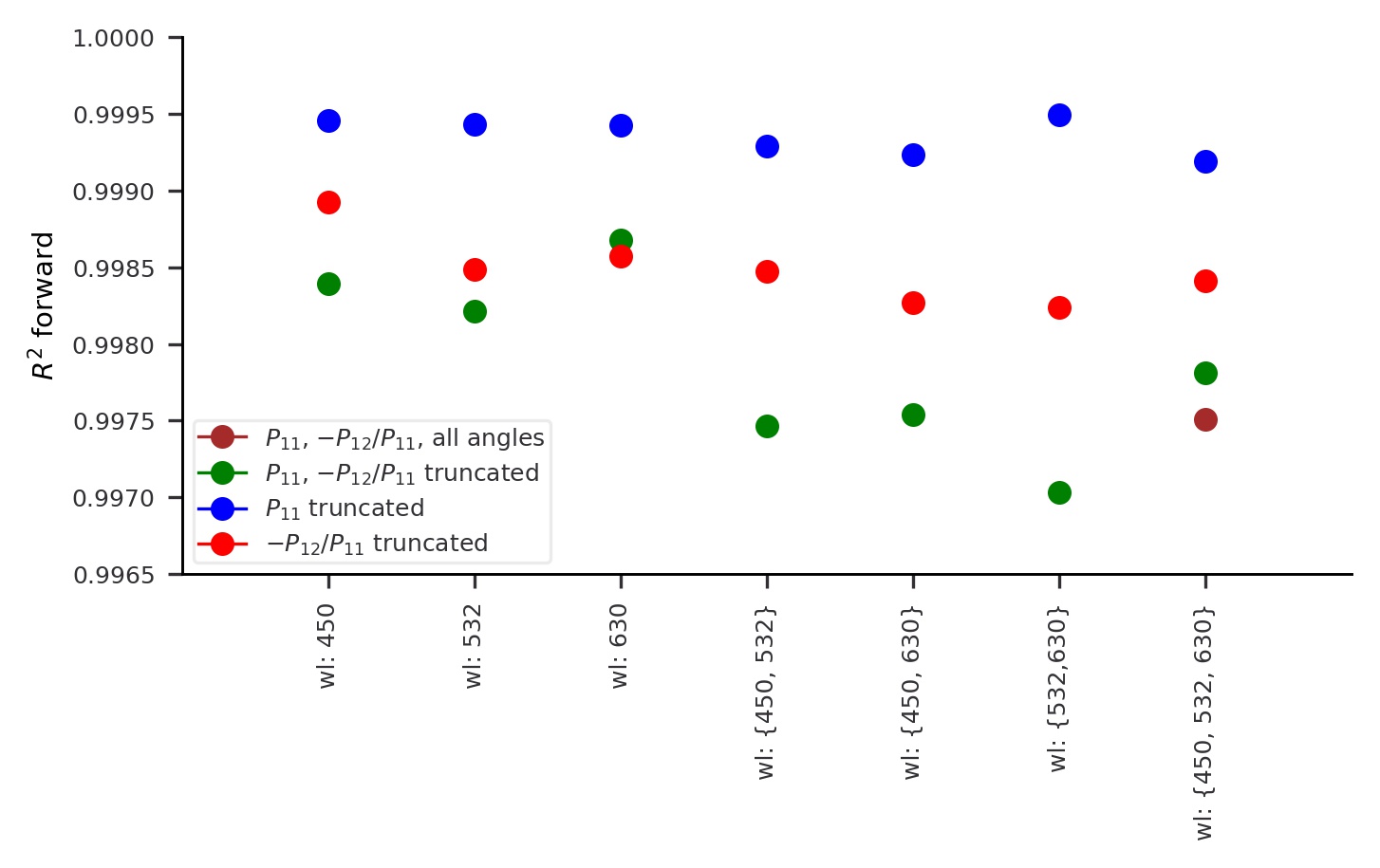}
    \caption{The $R^2$ values for the forward prediction for the different cases for the bimodal data set.  }
    
    \label{fig:r2_test_qoi_bim}
\end{figure}

In Figure \ref{fig:r2_test_qoi} and \ref{fig:r2_test_qoi_bim} the $R^2$ values for the monomodal and bimodal data set for the different networks in the case study are presented. 

\begin{figure}
    \centering
    \includegraphics[width=\textwidth]{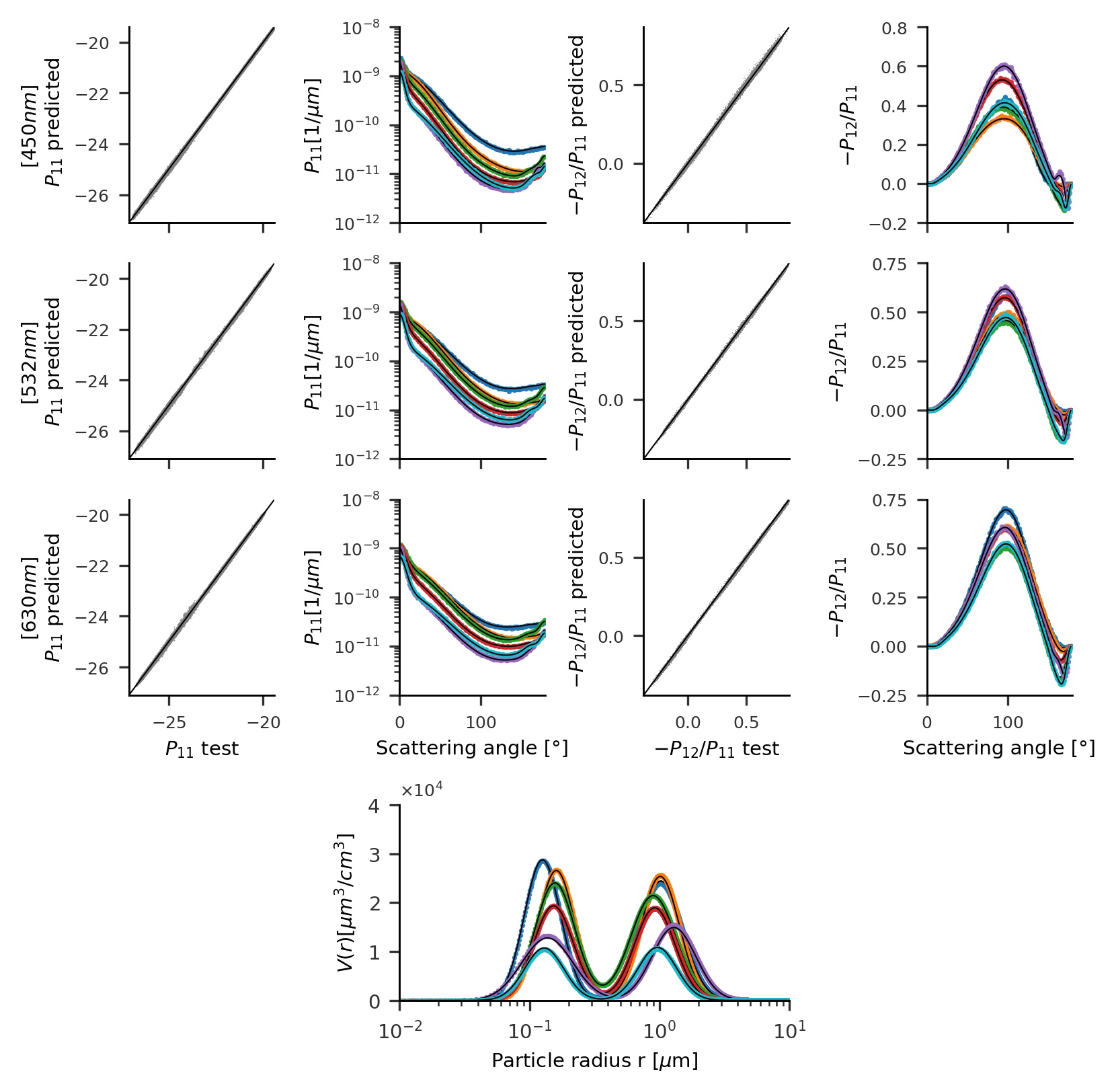}
    \caption{Comparison of the predicted and simulated test data for the bimodal case. First and third column shows correlation plots of the predicted versus the test data values of the (polarized) phase functions for all the three wavelengths. Second and fourth column depict the simulated (black line) versus predicted (colored dots) (polarized) phase functions in a qualitative representation. Six different data samples were chosen.}
    \label{fig:test_pred_bim}
\end{figure}

\begin{figure}
    \centering
    \includegraphics[width=\textwidth]{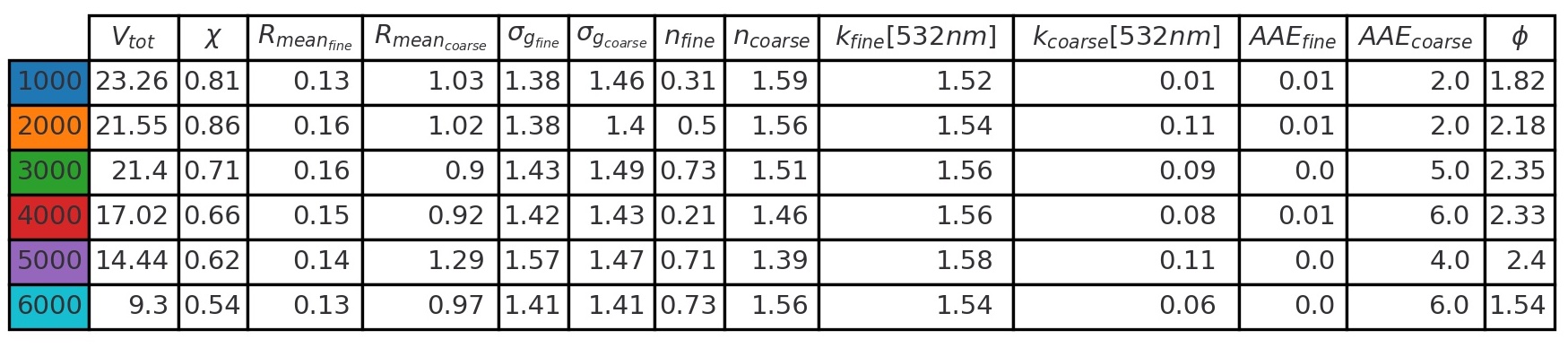}
    \caption{Aerosol property values corresponding to the selected samples in Figure \ref{fig:test_pred_bim}}
    \label{fig:pred_plot_values_bim}
\end{figure}

Figure \ref{fig:test_pred_bim} depicts the comparison between test data and predicted test data for the bimodal case. The corresponding aerosol property values for the six randomly chosen data samples can be found in Table \ref{fig:pred_plot_values_bim}.

\begin{figure}
    \centering
    \includegraphics[width=\textwidth]{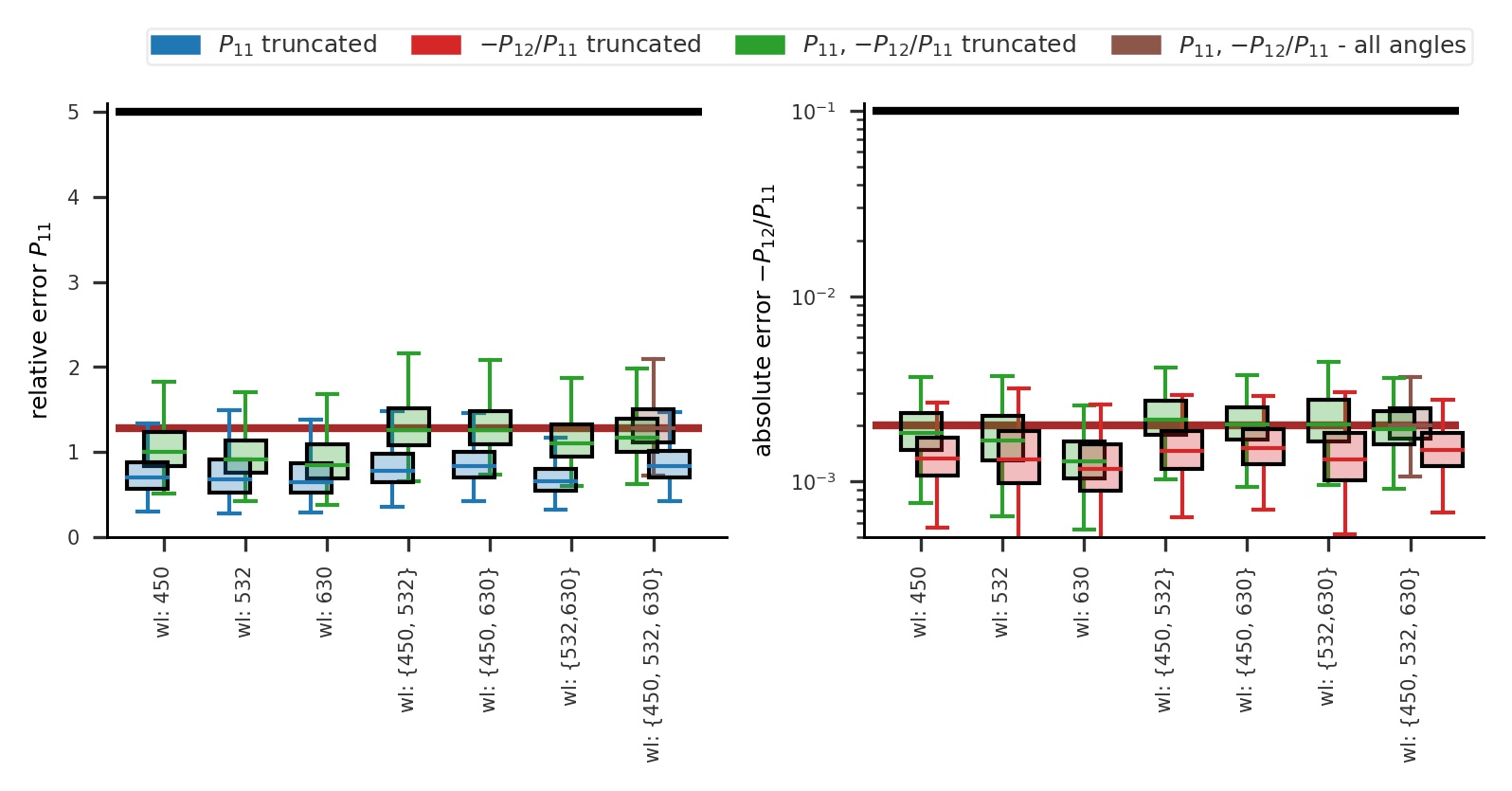}
    \caption{Box plots of the relative error of the phase functions and the absolute errors for the polarized phase functions for all 22 models for the bimodal case. Each boxplot includes a box that goes from the lower to the upper quartile values and a line at the median value of the absolute errors. The whiskers show the range of the absolute errors over the 20 000 data points in the test set. The brown solid line shows the median error of the full data case. The black solid line indicates the assumed measurement device errors.}
    \label{fig:abs_err_qoi_bim}
\end{figure}

The boxplots of the mean absolute errors for the forward model for the bimodal data set are depicted in Figure \ref{fig:abs_err_qoi_bim}.

%\begin{table}[]
%\begin{tabular}{lrrrrrr}
% &$ \phi [\%]$ & $n_{532}$ & $k_{532}$ x 100 &$R_{mean_{fine}} [nm]$ & $R_{mean_{coarse}} [nm]$ & $\chi$ \\ 
% RMSE literature& 35.9 & 0.044 & 1.34 & 9.94 & 163 & 0.029 \\
% RMSE INN & 0.7588 & 0.0021 & 0.11 & 0.25 & 4.34 & 0.03
%\end{tabular}
%\caption{Root Mean Square Error (RMSE) comparison between paper \ref{} and values for the test data set - for the RMSE of the refractive indices the mean of fine and coarse mode were taken for the INN}
%\label{tab:datasets_nr}
%\end{table}

%% If you have bibdatabase file and want bibtex to generate the
%% bibitems, please use
%%

 \bibliographystyle{elsarticle-num} 
 \bibliography{main.bib}

%% else use the following coding to input the bibitems directly in the
%% TeX file.

% \begin{thebibliography}{00}

% %% \bibitem{label}
% %% Text of bibliographic item

% \bibitem{}

% \end{thebibliography}
\end{document}